\documentclass[11pt]{article}
\usepackage{xspace,amsmath,amssymb}
\usepackage{color}
\usepackage{mathrsfs}
\usepackage[colorlinks=true,linkcolor=blue,filecolor=blue,citecolor=blue,urlcolor=blue]{hyperref}
\newcommand{\namedref}[2]{\hyperref[#2]{#1~\ref*{#2}}}
\renewcommand{\eqref}[1]{\hyperref[#1]{~(\ref*{#1})}}
\newtheorem{thm}{Theorem}[section]      % A counter for Theorems etc
\newtheorem{lemma}[thm]{Lemma}

\newtheorem{corollary}[thm]{Corollary}
\newtheorem{proposition}[thm]{Proposition}

\newcommand{\commentout}[1]{}
\newcommand{\IN}{\mbox{$I\!\!N$}}
\newcommand{\IR}{\mbox{$I\!\!R$}}

\newcounter{deff}
\newcommand{\bit}{\ensuremath{\{0,1\}}}

\newenvironment{sketch}{\noindent\emph{Proof Sketch:}}{\qed}
\newenvironment{proof}{\noindent\emph{Proof:}}{$\quad \Box$}

\newcommand{\hide}[1]{}
\def\qed{\vrule height7pt width4pt depth1pt\par}

\newenvironment{gproof}{\noindent{\bf Proof Sketch:~~}}{\qed}
\newcommand{\BPF}{\begin{gproof}} \newcommand {\EPF}{\end{gproof}}
\newenvironment{fproof}{\noindent{\bf Proof:~~}}{\qed}
\newcommand{\BPRF}{\begin{fproof}} \newcommand {\EPRF}{\end{fproof}}
\newenvironment{hproof}{\noindent{\bf Proof~}}{\qed}
\newcommand{\BPR}{\begin{hproof}} \newcommand {\EPR}{\end{hproof}}

\newcommand{\BI}{\begin{itemize}}
\newcommand{\EI}{\end{itemize}}
\newcommand{\BE}{\begin{enumerate}}
\newcommand{\EE}{\end{enumerate}}

\newcommand{\BT}{\begin{thm}}   \newcommand{\ET}{\end{thm}}
\newtheorem{dfn}[thm]{Definition}      %
\newcommand{\BD}{\begin{dfn}}   \newcommand{\ED}{\end{dfn}}
\newtheorem{corr}[thm]{Corollary}      %
\newcommand{\BCR}{\begin{corr}} \newcommand{\ECR}{\end{corr}}
\newtheorem{constr}[thm]{Construction}
\newcommand{\BCT}{\begin{constr}} \newcommand{\ECT}{\end{constr}}
\newtheorem{prop}[thm]{Proposition}
\newcommand{\BP}{\begin{prop}}   \newcommand{\EP}{\end{prop}}
\newtheorem{lemm}[thm]{Lemma}   % A counter for Lemmas etc
\newcommand{\BL}{\begin{lemm}}   \newcommand{\EL}{\end{lemm}}
\newtheorem{clm}[thm]{Claim}            %
\newcommand{\BCM}{\begin{clm}}   \newcommand{\ECM}{\end{clm}}
\newtheorem{sclm}[thm]{Sub-Claim}            %
\newcommand{\BSCM}{\begin{sclm}}   \newcommand{\ESCM}{\end{sclm}}
\newtheorem{assumption}[thm]{Assumption}            %
\newcommand{\BA}{\begin{assumption}}   \newcommand{\EA}{\end{assumption}}
\newtheorem{remark}[thm]{Remark}            %
\newcommand{\BR}{\begin{remark}}   \newcommand{\ER}{\end{ER}}
\newtheorem{example}[thm]{Example}

\newcommand{\complex}{\canon}
\newcommand{\complexity}{\complex}

\newcommand{\M}{{\cal M}}

\newcommand{\G}{{\cal G}}
\newcommand{\T}{{T}}

\newcommand{\C}{{\sf comm}}
\newcommand{\F}{{\cal F}}
\newcommand{\Z}{{\cal Z}}

\newcommand{\Exp}{{\mathbf{E}}}

\newcommand{\bitset}{\{0,1\}}

\newcommand{\N}{\mbox{N}}

\DeclareRobustCommand*{\slashfracstyle}[1]{%
  {\ensuremath{\mbox{\fontsize\sf@size\z@\selectfont #1}}}}

\DeclareRobustCommand*{\slashfrac}[2]{\leavevmode
  \raise.5ex\hbox{\scriptsize #1}\kern-.13em/%
  \kern-.15em\lower.25ex\hbox{\scriptsize #2}}

\def\*{\Z^*}

\def\B{\{0,1\}}
\def\B*{\B^*}

\def\union{\cup}

\newcommand{\view}{{\sf view}}

\newcommand{\canon}{\cC}

\def\cC{\mathscr{C}}

\def\cH{{\cal H}}

\def\cN{{\cal N}}

\def\cQ{{Q}}

\newcommand{\barw}{h}

\newcommand{\fullv}[1]{#1}
\newcommand{\shortv}[1]{\commentout{#1}}
\def\beginsmall#1{\vspace{-.5em}\begin{#1}\itemsep-\parskip}
\def\endsmall#1{\end{#1}\vspace{-1em}}
\newenvironment{RETHM}[2]{\trivlist \item[\hskip 10pt\hskip\labelsep{\bf
#1\hskip 5pt\relax\ref{#2}.}]\it}{\endtrivlist}
\newcommand{\rethm}[1]{\begin{RETHM}{Theorem}{#1}}
\newcommand{\repro}[1]{\begin{RETHM}{Proposition}{#1}}
\newcommand{\relem}[1]{\begin{RETHM}{Lemma}{#1}}
\newcommand{\recor}[1]{\begin{RETHM}{Corollary}{#1}}
\newcommand{\erethm}{\end{RETHM}}
\newcommand{\erepro}{\end{RETHM}}
\newcommand{\erelem}{\end{RETHM}}
\newcommand{\erecor}{\end{RETHM}}

\usepackage{chicagor}
\fullv{\bibliographystyle{chicagor}}
\shortv{\bibliographystyle{plain}}

\newcommand{\bott}{\omega}
\begin{document}
%raf95:
%\title{Game Theory with Costly Computation}
\title{Algorithmic Rationality: Game Theory with Costly Computation%
%joe104: moved up from below
\thanks{
A first draft of this paper appeared in April, 2007.  An extended
abstract with the title ``Game theory with costly computation'' appears
in \emph{First Innovations in Computer Science Conference}, 2010.
Halpern is supported in part by NSF grants ITR-0325453,
%joe108
%IIS-0534064, IIS-0812045, and IIS-0911036, and by AFOSR grants
%FA9550-08-1-0438 and FA9550-09-1-0266, and ARO grant %W911NF-09-1-0281.
IIS-0534064, IIS-0812045, and IIS-0911036, by AFOSR grants
%FA9550-08-1-0438 and FA9550-09-1-0266, and by ARO grant W911NF-09-1-0281.
Pass is supported in part by a Microsoft Research Faculty Fellowship, 
NSF CAREER Award CCF-0746990, AFOSR Award
FA9550-08-1-0197, and BSF Grant 2006317.}
%raf57: added affiliation
\author{
Joseph Y. Halpern\\
Cornell University\\
halpern@cs.cornell.edu 
\and Rafael Pass\\
Cornell University\\
rafael@cs.cornell.edu}
%raf62: fixed dates
%\date{First Draft: February 2007\\This Draft: July 2008}
%joe68: it's been a while.  Do we want to include this for FOCS?
%raf76: what do you think; i am ok with either options, should we also
%ref our CoRR report? 
%joe70: I don't feel strongly.  This is standard for economists.  My one
%concern with doing this for FOCS is that people will reject it for
%being ``old'' (although a lot of people already know how old it is.  So
%I have a very slight preference for not including it.
%\date{First Draft: April 2007\\This Draft: August 2008}
%raf77: removed then (but isn't it strange that we talk about subsequent
%work then?) 
%raf90:
%raf94
%\date{First Draft: April 2007\\This Draft: September 2009}
%raf101:
%\date{First Draft: April 2007\\This Draft: December 2009}
%joe104: cut for now
%\date{First Draft: April 2007\\This Draft: December
%2009
%\footnote{Minor updates in September 2011. An extended abstract
%  of this paper 
% appeared in the 1st Innovations in Computer Science Conference, 2010.
%First draft from April 2007.
}
%raf89:
%\date{September 14, 2009\footnote{First draft from April 2007.}}
\maketitle

\begin{abstract}
We develop a general game-theoretic framework for reasoning about
strategic agents performing possibly costly computation. 
In this framework, many traditional game-theoretic results (such as the
existence of a Nash equilibrium) no longer hold.  Nevertheless, we can
%joe91: 
%use the framework to provide psychologically appealing explanations to
use the framework to provide psychologically appealing explanations of
observed behavior in 
well-studied games (such as finitely repeated prisoner's dilemma and
rock-paper-scissors). 
Furthermore, we provide natural conditions on games sufficient to
guarantee that equilibria exist.

%joe106: added
\vspace{.2in}

%raf90: kill below
\iffalse
As an application of this framework, we
%consider a notion of game-theoretic implementation of 
develop a definition of protocol security relying on
game-theoretic notions of implementation. % of mediators.  
%mediators in computational games. 
We show
that a natural special case of this
%this notion
this definition
is equivalent to a variant of 
the traditional cryptographic definition of protocol security;
%\emph{precise zero-knowledge}
%(Micali and Pass, STOC'06).
this result 
shows that, when taking computation into account,
the two approaches used for dealing with ``deviating'' players
in two different communities---\emph{Nash equilibrium} in game theory
and \emph{zero-knowledge ``simulation''} in cryptography---are intimately
related.

%raf82: cutting as per joe73
%Other special cases of our definition instead 
%lead to more practical protocols and circumvent known lower bounds
%with respect to the cryptographic notion of security.
%raf81: should we cut some of the crypto stuff here?
%joe73: we could perhaps cut the last sentence.
%raf82: ok.
\fi
\end{abstract}
\thispagestyle{empty}
\newpage
\pagenumbering{arabic}
%raf58: put this back, but am happy to remove it too
%joe57: I think we should drop it if we split the paper; it's somewhat
%nonstandard, and perhaps not so necessary for a shorter paper, but we
%can leave it for this version. 

%\tableofcontents
\newpage

\section{Introduction}
%raf90: removing old commented out stuff
Consider the following game. You are given a random odd $n$-bit number $x$ and you are supposed
to decide whether $x$ is prime or composite. If you guess correctly you receive \$2, if you guess incorrectly you instead have to pay a penalty of $\$1000$. Additionally you have the choice of ``playing safe'' by giving up, in which case you receive $\$1$.
In traditional
game theory, computation is considered ``costless''; in other words,
players are allowed to perform an unbounded amount of computation
without it affecting  
their utility. Traditional game theory suggests that you should compute
whether $x$ is prime or composite and output the correct answer; this
is the only Nash equilibrium of the one-person game, no matter what $n$
(the size of the prime) is.
Although for small $n$ this
seems reasonable, when $n$ grows larger most people would probably
%joe3
%decide to ``play safe''---as eventually the cost of computing the answer 
decide to ``play safe''; eventually the cost of computing the answer
(e.g., by buying powerful enough computers) outweighs the possible gain of
$\$1$. 

%raf107:
%The importance of considering such computational issues in game theory
The importance of considering such computational issues in game theory
has been recognized since at least the  work of Simon \citeyear{Simon55}.
There have been a number of attempts to capture various aspects of
computation. Two major lines of research can be identified.
The first line, initiated by Neyman \citeyear{Ney85}, tries to model
the fact that players can do only bounded computation, typically by
modeling players as finite automata.  
%raf94: copied from ics
%joe85: % missing in the next round, which I added
%joe80: added next line, to make it more of a story
%the flow, and it's confusing because now both MW and DHR consider TMs
Neyman focused on finitely repeated prisoner's dilemma, a topic which
%joe90
%has contined to attract attention. 
has continued to attract attention. 
%joe80
(See \cite{PY94} and the references
%joe21: shaving
%therein for more recent work on this approach.)  The second line, initiated
%therein for more recent work on this approach.)  
%raf92:
%therein for more recent work on modeling players as finite automata.
%Megiddo and Wigderson \cite{MW} instead model players as Turing
%machines; instead model players in prisoner's dilemma as Turing machines
therein for more recent work on prisoner's dilemma played by finite automata;
Megiddo and Wigderson \citeyear{MeWi} considered 
prisoner's dilemma played by Turing machines.)
%joe80*: This is a different problem: implementing mediators.  I rewrote
%it and corrected typos. 
%a more recent approach, first considered by Urbano and Villa
%\cite{UV04}
%and formalized by Dodis, Halevi and Rabin \cite{DHR00}, instead models
%players as polynomially bounded Turing machine.) 
In another instance of this line of research, Dodis, Halevi, and Rabin
\citeyear{DHR00} and Urbano and Vila \citeyear{UV04} consider the
problem of implementing mediators when players are polynomial-time
Turing machines.

%raf102: this was repeated
%(See \cite{PY94} and the references
%therein for more recent work on modeling players as finite automata;
%a more recent approach, first considered by Urbano and Villa
%\citeyear{UV04}
%and formalized by Dodis, Halevi and Rabin \citeyear{DHR00}, instead models
%players as polynomially bounded Turing machine.) 
The second line, initiated
by Rubinstein \citeyear{Rub85}, tries to capture the fact that doing
%joe8: typo
%costly computation affects the agents utility.  In particular, 
%joe21
%costly computation affects an agent's utility.  In particular, 
costly computation affects an agent's utility. 
%raf60:
%joe59: I'm not sure what changes this raf60 is referring to, or the
%next one.  But everything looks fine to me in any case.
%raf61: sorry, i did some changes but then removed them
Rubinstein assumed that players choose a finite automaton to play the 
game rather than choosing a strategy directly; a player's utility
depends both  
on the move made by the automaton and the complexity of the automaton
(identified with the number of states of the automaton).
Intuitively, automata that use more states are seen as representing more
complicated procedures.  
(See \cite{Kalai90} for an overview of the work in
this area in the 1980s, and \cite{BKK07} for more recent work.)
%raf107* added. not sure if this is the right place. also, i have a
%hard time seeing where in the original sandholm paper this is done,
%but the later LS paper reference him so, i am fine leaving it in.
%the papers below are the first, second, third and last on the
%reviewers list.
%joe104: ``recent'' is a moving target ...
Another growing literature (see, e.g.,
%raf108: hi joe, why did you change the order? it was in chronological
%order (and we use the same format in the sentence about regardin more
%recent work)
%joe105: The trouble with chronological order is that it gets tricky
%with regard to journal references rather than conference references, so
%I've always used alphabetical order, although I don't feel strongly
%about it.
%\cite{Sandholm,LarsonSandholm1,LarsnSandholm2}, and \cite{CKLNP12}
%for more recent work) consider the effict of 
%costly delibaration in auctions, where an agent is uncertain about its
%preference and can take costly actions (e.g., computation) to reduce
%its uncertainty.
%joe104: Rafael, can you send me these references?
\cite{CKLNT12,LarsonSandholm1,LarsonSandholm2,Sandholm00})
considers the effect of
costly delibaration in auctions, where an agent is uncertain about her
preferences and can take costly actions (e.g., computation) to reduce
her uncertainty.

Our focus is on providing a general game-theoretic framework for reasoning
about agents performing costly computation.   
As in Rubinstein's work, we view players as choosing a machine, but for
us the machine is a Turing machine, rather than a finite automaton.  We
associate a complexity, not just with a machine, but with the machine
and its input.
The complexity 
could represent the running time of or space used by the machine on that
input.  The complexity can also be used to capture the complexity of the
machine itself (e.g., the number of states, as in Rubinstein's case) or to
model the cost of searching for a new strategy to replace one that the
player already has.  For example, if a mechanism designer recommends
that player $i$ use a particular strategy (machine) $M$, then there is a
cost for searching for a better strategy; switching to another strategy
may also entail a psychological cost.  By allowing the complexity to
depend on the machine \emph{and} the input, we can deal with the fact 
that machines run much longer on  some inputs than on others. 
A player's utility depends both
on the actions chosen by all the players' machines and the complexity of
these machines.  

%raf94: remove as per joe79
%Note that, in general, unlike earlier papers, player
%$i$'s utility may depend not just on the complexity of $i$'s machine,
%but also on the complexity of the machines of other players.  For
%example, it may be important to player 1 to compute an answer to a
%problem before player 2.
%raf30: added back more emphasis on input
%joe21: moved up
%Additionally, unlike earlier works,
%the complexity of a machine depends
%not only on the machine, but also its input; this distinction becomes
%important when considering machines that 
%%joe21
%%upon receiving certain inputs run much longer than on others. 
%run much longer on  some inputs than on others. 
%joe8*: done.  I talked  (although I didn't make a fuss here about what
%the input to 
%a machine is)
%\Rnote{Furthermore and most importantly, unlike earlier paper which assign a 
%complexity each machine, we assign a complexity to a machine based on the 
%{actual computation} it has performed; in particular the complexity is a
%function  
%that depends on the type of a player, its random coins (and potentially
%also the messages it receives).}

In this setting, we can define Nash equilibrium in the obvious way. 
However, as we show by a simple example (a rock-paper-scissors game 
%joe56
%with minimal assumptions about the complexity of machines), a Nash
where randomization is costly), a Nash
equilibrium 
may not always exist.  
%raf50:
%On the other hand, by choosing complexity
%appropriately, we can make any tuple of machines a Nash equilibrium.
%raf56: removed to put back old
%Other standard results in the game theory also do not hold.
%joe6*: 
%raf50:
%raf56: put back
%raf102
%Other standard results in the game theory, such as the
Other standard results in game theory, such as the
\emph{revelation principle} (which, roughly speaking, says that there is always an
equilibrium where players truthfully report their types, i.e., their
%raf56: put Myerson, Forges instead of OR
%private information \cite{OR94}) also do not hold .
%joe105: alphabetized
%private information \cite{Myerson79,F86}) also do not hold. 
private information \cite{F86,Myerson79}) also do not hold. 
We view this as a feature.  We believe that taking computation
into account should force us to rethink a number of basic notions.  
%raf50:
%joe33
%In particular, we introduce refinements of Nash Equilibria that
%consider
%joe56
%In particular, 
%raf81: this is now pushed into the crypto part.
%raf90: put this back now, since we do have seq eq
%raf107: oops this is no longer in teh paper
%To this end,
%we introduce refinements of Nash equilibrium that take
%into account the
%computational aspects of games.
%joe56: rewrote material from below
Moreover, as we show here by example, despite the fact that Nash
equilibrium may not always exist, there are interesting Nash equilibria
that give insight into a number of games. 

%raf81:
First, we show that the non-existence of Nash equilibrium is not such a
significant problem.  
%raf81: ``newish'' below
We identify natural conditions (such as the assumption that
randomizing is free) that guarantee the existence of Nash equilibrium in
computational games.
Moreover, we demonstrate that our computational framework can 
%joe73
%explain experimentally-observed phenomena (e.g., in the 
explain experimentally-observed phenomena, such as cooperation in the
%raf81: added but then removed the other examples
%---such as usage of \emph{tit-for-tat} in 
finitely repeated prisoner's dilemma,
%, \emph{biases in information processing}, or usage of \emph{weak
%randomization}--- 
%joe73
%which are inconsistent with classical game-theoretic models in a
that are inconsistent with standard Nash equilibrium in a
psychologically appealing way. 
%raf85: fixed ref inside
Indeed, as shown by our results, many concerns expressed by the emerging
field of \emph{behavioral economics} (pioneered by Kahneman and Tversky
%joe76
%\cite{kt81}) can be accounted for by simple assumptions about players'
\citeyear{kt81}) can be accounted for by simple assumptions about players'
cost of computation,  
%raf81: is ``ad-hoc'' better than ``complicated''?
%joe73: I think so
%without resorting to complicated cogntivite or psychological models.
without resorting to ad hoc cognitive or psychological models.
%raf81: old below
%raf107*:
We emphasize that it is not the basic framework itself that
%joe104
%restricts behavior; \footnote{The only restriction imposed by our
%framework is that player strategies are computable.} rather,
restricts behavior;\footnote{The only restriction imposed by our
  framework is that players' strategies are computable.} rather,
%joe104
%behavior is restricted by considering specific (but in our eyes
%natura) complexity costs (e.g., that randomness is costly, or memory is
%costly). 
behavior is restricted by considering specific (but arguably
natural) complexity costs (e.g., charging for randomization or memory).

%raf81:
%raf90:
Finally, we
%We also 
show that our framework is normative in that it can be used
%joe73: slight change
%to tell a mechnism designer how to design a mechanism. 
%joe90
%to tell a mechnism designer how to design a mechanism that takes into
to tell a mechanism designer how to design a mechanism that takes into
account agents' computational limitations.
We illustrate this point in the context of cryptographic protocols.
% raf90
In a companion paper \cite{HP09b}, we use our framework to provide  
%We use our framework to provide 
%Finally, as an application of this framework we provide 
a game-theoretic
definition of 
%raf90
cryptographic protocol security.
\paragraph{Paper Outline}
The rest of this paper is organized as follows.  In
%raf98:
%Section~\ref{sec:framework} we describe our framework.  
%joe87
%Section~\ref{sec:framework} and \ref{sec:mediator} we describe our
%framework.  
Sections~\ref{sec:framework} and \ref{sec:mediator} we describe our
framework.   
%raf90:
%joe104
%In Section \ref{sec:behavior} show how our computational framework
In Section \ref{sec:behavior}, we show how our computational framework
can provide ``psychologically-appealing'' explanations to observed behavior that is inconsistent with traditional solution concepts.
Section \ref{sec:suff} contains our existence results for Nash equilibrium.
%joe101: cut, 
%In Section \ref{sec:seq} we provide definitions and existence results 
%for sequential equilibria. 
We conclude in Section~\ref{sec:conclusion} with
%joe103
%potential 
new directions of research made possible by our framework.

\section{A Computational Game-Theoretic Framework}\label{sec:framework}
\subsection{Bayesian Games}\label{sec:Bayesian games}
We model costly computation using \emph{Bayesian machine games}.  To
explain our approach, we first
review the standard notion of a \emph{Bayesian game}.
A Bayesian game is a
game of incomplete information, where each player makes a single move.
The ``incomplete information'' is captured by assuming that nature makes
an initial move, and chooses for each player $i$ a \emph{type}
in some set $T_i$.  Player $i$'s type can be viewed as describing $i$'s
private information.  For ease of exposition, we assume in this paper
that the set $N$ of players is always $[m] = \{1,\ldots, m\}$, for some $m$.
If $N = [m]$, the set $T =
T_1 \times \ldots \times T_m$ is the \emph{type space}.  
%joe13
As is standard, 
we assume that
there is a commonly-known 
%joe13
%distribution $\Pr$ over the type space T.  Each player $i$ 
probability distribution $\Pr$ on the type space $T$.  Each player $i$ 
must choose an action from a space $A_i$ of actions.  Let $A = A_1
%joe99
%\times \ldots \times A_n$
\times \ldots \times A_m$
be the set of action profiles.  A Bayesian game is
characterized by the tuple $([m],T,A,\Pr, \vec{u})$, where $[m]$ is the
set of players, $T$ is the type space, $A$ is the set of joint actions,
and $\vec{u}$ is the utility function, where $u_i(\vec{t},\vec{a})$ is
player $i$'s utility (or payoff) if the type profile is $\vec{t}$ and
action profile $\vec{a}$ is played.  

In general, a player's choice of action will depend on his type.
A \emph{strategy} for player $i$ is a function from $T_i$ to
$\Delta(A_i)$ (where, as usual, we denote by $\Delta(X)$ the set of
distributions on the set $X$).  If $\sigma$ is a strategy for player
%joe103
%$i$, $t \in T_i$ and $a \in A_i$, then  $\sigma(t)(a)$ denotes the 
$i$, $t \in T_i$, and $a \in A_i$, then  $\sigma(t)(a)$ denotes the 
probability of action $a$ according to the distribution on acts
induced by $\sigma(t)$.  Given a joint strategy
$\vec{\sigma}$, we can take $u_i^{\vec{\sigma}}$ to be the random
variable on the type space $T$ defined by taking
$u_i^{\vec{\sigma}}(\vec{t}) = \sum_{\vec{a} \in A}
(\sigma_1(t_1)(a_1) \times \ldots \times \sigma_m(t_m)(a_m))u_i(\vec{t},
\vec{a})$. Player $i$'s 
%raf107*: added 
%(ex-ante)
(ex ante)
expected utility if $\vec{\sigma}$ is
played, denoted $U_i(\vec{\sigma})$, 
is then just $\Exp_{\Pr}[u_i^{\vec{\sigma}}] 
%joe12
%that is, $i$'s expected utility if $\vec{\sigma}$ is played, taken with
%joe17: why not just write the definition
%where the expectation is taken with
%respect to the probability $\Pr$ on types. 
= \sum_{\vec{t} \in T} \Pr(\vec{t})u_i^{\vec{\sigma}}(\vec{t})$.
%raf107*: added for AE
%We may also consider player $i$'s expected utility conditoned on player
%joe108
%We can also consider player $i$'s expected utility conditoned on player
We can also consider player $i$'s expected utility conditioned on player
$i$'s type $t_i$: $U_i(\vec{\sigma} \mid t_i) = \Exp_{\Pr'}[u_i^{\vec{\sigma}}] 
= \sum_{\vec{t}_{-i} \in T_{-i}}
\Pr'(\vec{t})u_i^{\vec{\sigma}}(\vec{t})$, where $\Pr'$ is $\Pr$
conditioned on player $i$'s type being $t_i$.

\fullv{\subsection{Bayesian Machine Games}}
%raf72:
%\nshortv{\paragraph{Bayesian machine games}}
%A \emph{Bayesian machine game} is similar in spirit to the standard
%notion of a \emph{Bayesian game} in game theory.  (We review Bayesian
%games in the appendix, for the reader's convenience.)
In a Bayesian game, it is implicitly assumed that computing a
strategy---that is, computing what move to make given a type---is free.  
We want to take the cost of computation into account here.  
\fullv{To this end,
we consider what we call \emph{Bayesian machine games}, where we
replace strategies by \emph{machines}.}
\shortv{In a Bayesian machine game, we replace strategies by machines.}
%joe10
%A machine
%could be, for example, a Turing machine, a program, or a
%finite state automaton; 
For definiteness, we take the machines to be Turing machines, although 
the exact choice of computing formalism is not
significant for our purposes.  
\iffalse
What does matter is that we can
talk about \emph{complexity} of a machine,
where complexity for instance
can be thought of as 
the number of steps taken by the machine and the size of 
(the representation of) the
machine, where the size can be thought of as the number of states
in the Turing machine or finite automaton, or the length of the program,
if we think of a machine as a program in a programming language.
\fi
%joe7: cut for abstract
\fullv{
Given a type, a strategy in a Bayesian game returns a distribution over
actions.  Similarly, given as input a type, the machine returns a
distribution over actions.  
%raf94: remove para below as per joe73? (since you repeat it more
%succintly below)
%joe85: yes, I agree (so I removed it) 
%As is standard, we model the distribution by
%assuming that the machine actually gets as input not only the type, but
%a random string of 0s and 1s (which can be thought of as the sequence of
%heads and tails), and then (deterministically) outputs an action.
}%
Just as we talk about the 
%joe12
expected
utility of a strategy profile in a Bayesian
game, we can talk about the expected utility of a machine profile in a
Bayesian machine game.  However, we can no longer compute the
expected utility by just taking the expectation over the action profiles
that result from playing the game.  
%Bayesian machine games can be viewed as a special case of Bayesian games
%that allow us to take computational considerations into account.  
%In a Bayesian machine game, rather than choosing a strategy, a player
%chooses a (Turing) machine to implement a strategy.  Just as with a
%strategy, the action performed by the machine depends on the player's
%type.  However, 
A player's utility depends not only on the type profile
and action profile played by the machine, but also on the
%joe4
%complexity and number of computation steps taken by the machines
%chosen,
``complexity'' of the machine 
%joe10: added
given an input.
%and the cost of  
%``implementing'' the machines
%joe4*
The complexity of a machine can represent, for example, the 
%joe10
%number of computation steps, the space usage of the machine, 
running time or space usage of the machine on that input, the size
of the
program description, or  some combination of these factors.  For
simplicity, we describe the complexity by a single number, although,
since a number of factors may be relevant, it may be more appropriate to
represent it by a tuple of numbers in some cases.  
%joe21
\fullv{
(We can, of course,
always encode the tuple as a single number, 
but in that case, ``higher''
complexity is not necessarily worse.)  
}
%joe10
%Of course, features such as the running time of the algorithm may
%depend on the input to the machine.  
Note that when determining player $i$'s
%joe4
%utility, we consider the complexity and number of steps taken by 
utility, we consider the complexity of 
all machines in the profile, not just that of $i$'s machine.  
For example, $i$ might be happy as long as his
%joe13
%machine takes fewer steps than that of $j$'s.  
machine takes fewer steps than $j$'s.  
%joe6*: 
%raf30: removed
%While we place essentially no constraints on the form of the complexity
%function, we do have the intuition that the only machines that can have a
%complexity of 0 are those that do absolutely nothing.  (This intuition
%is formalized in Definition~\ref{def:natural}.)
%The intuition behind implementation is that if a 
%particular machine is recommended as part of a mechanism, while there
%may be a better choice out there, finding and implementing that better
%choice also entails a cost.   

%\Jnote{We should put in here some discussion of other
%approaches--Rubinstein, Kalai \& Kalai, etc.}

%For technical reasons, we make a simplifying assumption:
%\beginsmall{itemize}
%\item There exists some $m$ such that the type space of each player is
%$\{0,1\}^m$.  The reason for this is that we will be interested in
%families of games, parameterized by $m$, which are essentially the same
%except that the type space.
%%joe6*:
%
%raf107:
%We assume that nature has a type in $\{0,1\}^*$.  While there is no
We assume that nature has a type in $\{0,1\}^* = \cup_{k \in N} \{0,1\}^k$.  While there is no
need to include a type for nature in standard Bayesian games (we can
effectively incorporate nature's type into the type of the players), once
we take computation into account, we obtain a more expressive class of
%raf94:
%games by allowing nature to have a type.
games by allowing nature to have a type
%joe21*
%joe103*
%(since the complexity of computing the utility may depend on nature's type).
(since the complexity of a machine on a given input may depend on nature's type).
%\item  
%
We assume that machines take as input strings of 0s and 1s and 
output strings of 0s and 1s.  
%joe4
Thus, we assume that both types and actions can be represented as
%raf47:
%elements of $\{0,1\}^*$,
elements of $\{0,1\}^*$.
%raf47: back to terminating machine
%joe2: changed to terminating machines
%player's action can be represented by such a string.  We assume that
%all computations terminate, so the output is an element of $\{0,1\}^*$.  
%joe13*
%raf47:added footnote
%\footnote{This assumption is made for simplicity of presentation. Our results hold also when considering machines that terminate with probability 1.}
%raf56**: I did not change to consider machine that terminate with
%probability 1. Should we do this by adding a special output symbol for
%non terminating computation? 
%joe37: Hmm ... the usual symbol is \bot, but we've already used \bot
%for the TM that does nothing.  I'd actually prefer to use \bot for a
%nonterminating compuatation, and some other symbol (M_0?) for a TM that does
%nothing.  
%terminate, so the output is a finite string.\footnote{Our definitions
%make perfect sense even if machines sometimes do not terminate, although
%we then have to assign a utility to that situation. Our main technical
%results depend on having machines that always terminate (or, at least
%terminate with probability 1).}  %while output can be represented as
				 %elements of $\{0,1\}^\infty$ (so that 
%machines might not terminate, although in the cases of interest, this
%happens with probability 0).
We allow machines to randomize, so given a type as input, we actually
get a distribution over strings.  
%Since the machines are randomized, player $i$'s machines output may
%depend on player $i$'s type and a string chosen with uniform
%probability from 
%joe5
To capture this, 
we assume that the input to a machine is not only a type, but 
also a string chosen with uniform probability from 
$\{0,1\}^\infty$ (which we can view as the outcome of an infinite
%joe5
%sequence of coin tosses); we denote by $M(t;r)$ the output of machine
sequence of coin tosses).   
%raf94, joe79: added
The machine's output is then a deterministic function of its type and
the infinite random string.
%raf57: added
%joe56
%Implicit in the above representation is the assumption that machines 
%raf83:
%Implicit in the representation above is the assumption that machines 
%joe
%Implicit in the representation above is the convention that the output
%of machines that do not terminate is taken to be some special symbol
%$\bot$.
%raf95: don't need this in this paper
%raf98*: i removed the wrong para, put it back
%\iffalse
We use the convention that the output
of a machine that does not terminate is a fixed
%raf94 joe79: \bot is overloaded; I replaced it by \bott here, which I defined
%as \omega for now, but we can change it.
%special symbol $\bot$ .
%raf92:
%special symbol $\bott$ .
special symbol $\bott$.
%\fi
%assumption that machines 
%always terminate, so the output is a finite string.\footnote{This
%assumption is made for simplicity of presentation. Our results hold
%also when considering machines that terminate with probability 1.}  
%raf83: remove
%terminate with probability 1, so the output is a finite
%raf81: joe, couldn't we just take the output to be some special symbol
%(and the same for complexity) if the 
%machine does not abort? i know we must have discussed this but i can't
%joe73: I'm not quite sure what you have in mind here, Rafael.  If the
%machine runs forever, we can't have it's output being a special
%symbol.  Let's talk about this when we meet.
%remember why we decided against it. 
%raf83: remove footnote too
%string.\footnote{For ease of presentation, our notation ignores the
%joe56
%probability 0 events where a machine does not terminate. Technically,
%we need to assign a utility also to such events, but as long as this
%utility is finite these events can be ignored in our calculations.} 
%raf83: remove rest of footnote
\iffalse
possibility that a machine does not terminate. Technically, we
also need to assign a utility to inputs where this happens, 
but since it happens with probability 0, as long as this 
utility is finite, these outcomes can be ignored in our calculations.} 
\fi
%joe6
We define a \emph{view} to be a pair $(t,r)$ 
%joe13: redundundant
%to be a pair 
%joe7*
%consisting of a type and a random string.  We denote by $t;r$ a string in
%$\{0,1\}^\infty$ representing the view, under some suitable encoding.
of two 
%raf83: no longer finite
%\emph{finite} 
bitstrings; we think of $t$ as that part of the
%joe103
%type that is read, and $r$ is the string of random bits used. 
type that is read, and of $r$ as the string of random bits used. 
%raf32: added below, but it is kind of unclear what traditionally refers to.
%joe21
%(We mention that our definition is slightly different from the
(Our definition is slightly different from the
traditional way of defining a view, in that we include only the parts of
%joe21
%the type and the part of the random sequence \emph{actually} read. 
the type and the random sequence \emph{actually} read. 
If computation is not taken into account, there is no loss in generality
in including the full type and the full random sequence, and this is
what has traditionally been done in the literature.
%joe103
%However, when computation is costly, this might no longer be
However, when computation is costly, this may no longer be
the case.) 
We denote by $t;r$ a string in
%joe17*: 
%$\{0,1\}^*$ representing the view, under some suitable encoding.
$\{0,1\}^*; \{0,1\}^*$ representing the view.  
%raf94: added paren as per joe 79
(Note that here and
elsewhere, we use ``;'' as a special symbol that acts as a separator
between strings in $\bit^*$.)
%raf83*: is this really needed, also it becomes a bit funny now given
%that r might be infinite. 
%joe75: I reinstated this, just because we seem to use it in various
%places (and I did say above that the output of M when it doesn't halt
%is \bot, so it seems consistent
%raf85: but if r is already an infinite string you are now adding to
%infinite strings? 
%joe76: Sorry; I missed teh point
%If $v = (t;r)$, we take $M(v)$ to be the output of $M$ given input type
If $v = (t;r)$ and $r$ is a finite string, we take $M(v)$ to be the
output of $M$ given input type 
$t$ and random string $r\cdot 0^\infty$.

We use a \emph{complexity function} $\complex: {\bf M}\times \bitset^*
%joe21: a view is now more complicated
%raf40**: joe, i would like to ignore the ; problem, we no longer need
%it to be a special character, it is only used for ease, adding it here
%will complicate things when we consider mediated games... 
%joe22*: having ; certainly makes life much easier in a few cases.  If
%we say it's just an encoding (which is what we used to say), I think that
%would cause a number of small changes.  Maybe we can just take V to
%consist of all strings of the form x;y, where x,y are bitstrings?  That
%might solve the notational problem.
%raf41: sure!
%raf83: adde infty
%; \bitset^*
%joe103
%; \bitset^* \union \bitset^{\infty}
; (\bitset^* \union \bitset^{\infty})
%\rightarrow \IN$, where ${\bf M}$ denotes the set of Turing Machines, to
%raf57:
%raf72:
%\rightarrow \IN$, where ${\bf M}$ denotes the set of Turing Machines
%joe103
%\rightarrow \IN$, where ${\bf M}$ denotes the set of Turing machines
\rightarrow \IN$, where ${\bf M}$ denotes the set of Turing machines,
%joe56
%(which terminate with probability 1), to 
%raf83:
%that terminate with probability 1, to 
to
%joe21:
%describe the complexity of a given machine given a view. 
describe the complexity of a machine given a view. 
%raf83: moved this to crypto section
\iffalse
Given our intuition that the only machines that can have a complexity of
0 are those that ``do nothing'', we require 
%joe20*: tried to clarify
%a complexity function
%$\complex$ to assign positive complexity to all machines except for the
%(canonical) machine $\bot$ that does nothing; furthermore
%$\complex(\bot,\cdot) = 0$. 
that, for all complexity functions $\complex$, $\complex(M,v) = 0$ 
%joe21
%for some $v \in \bit^*$ iff $M %= \bot$ iff $\complex(M,v) = 0$ for all
%$v \in\bit^*$, 
for some view $v$ iff $M = \bot$ iff $\complex(M,v) = 0$ for all views $v$, 
where $\bot$ is a canonical represent of the TM that does
nothing: it does not read its input, has no state changes, and writes
nothing.
\fi  
%joe30*: to deal with your concern below
If $t \in \{0,1\}^*$ and $r \in \{0,1\}^\infty$, we identify 
$\complex(M, t;r)$ with $\complex(M,t;r')$, where $r'$ is the finite
prefix of $r$ actually used by $M$ when running on input $t$ with random
string $r$.
%joe6: moved back
%joe7*: may want to cut this ...
%Although we will restrict the types of agents in any particular game 
%to be in $\{0,1\}^m$, we require that machines be defined for all
%possible types in $\{0,1\}^*$.  This is useful when dealing with
%complexity issues, since it allows us to think of the player as using
%the same machine in a family of games.
%rafael
%joe5*: Just a small technical quibble.  Why do you take the view to be
%an infinite bitstring t;r rather than a pair (t,r)?  If you take the
%former view, you need a ``separator'' of some sort between t and r.
%Moreover, we should probably write t\cdot r (with no parens) if it's
%really a bitstring.  I left it as (t;r) for now, but called it a pair.
%The \emph{view} of a machine is an infinite bitstring string $v = (t;
%r)$, where $t$ is of finite length. To simplify notation we denote by
%$M(v)$, where  $v = t; r$, the output of  
%of $M$ on input the type $t$ and random string $r$ (i.e., $M(t;r)$).
%We also consider views $v'=(t';r')$ of \emph{finite} length. We abuse
%of notation and let $M(v')$ denote the output of the machine $M$ on
%input the type $t'$ and the random string $r'$ concatenated with
%$0^\infty$ (i.e., $M(t';r||0^\infty)$). 
%joe6: cut
%We call the pair $(t;r)$ the \emph{view} of a machine.  
%joe6: added next line
%We typically care not about the whole random string in a view, but just some
%finite prefix of it.
%\endsmall{itemize}

%raf95*: should we keep this here? now this is claimed as a contribution
%in gtsec.  
%joe85*: I think it will depend on how we deal with sequential
%equilibrium; see my email.
%joe9
\fullv{
For now, we assume that machines run in isolation, so the output and
%joe4
%running time of a machine does not depend on the machine profile.  We
complexity of a machine does not depend on the machine profile.  We
%raf94: joe79
%remove this restriction in the next section.
remove this restriction in the next section, where we allow machines to
communicate with mediators (and thus, implicitly, with each other via
the mediator).
}

%joe6: done above
%\Rnote{Need to add type for nature. 
%Is this standard? We need to add a discussion about this (and change the
%text above)} 
%\Jnote{If the results that we actually present in the abstract use the
%type, then I can easily add it; otherwise, it's probably not worth it.}
\BD [Bayesian machine game]
\label{bmg.def}
%joe4
%A Bayesian Machine Game $G$ is described by a tuple 
A \emph{Bayesian machine game $G$} is described by a tuple 
%joe2: removed subscripts on \timec, \size, and \switch
%$([m], m, \M,T, \Pr, \timec_1, \ldots, \timec_n, \size_1, \ldots, \size_n,
%\switch_1, \ldots, \switch_n, 
%joe4
%$([m], m, \M,\Pr, \timec, \size,
%joe6: broke line
%$([m], m, \M,\Pr,\complexity_1, \ldots, \complexity_n,
%$([m], m, \M,\Pr,$ $\complexity_1, \ldots, \complexity_n,
%raf18 removed type length
%joe56: need to reinsert the T!
%$([m], \M,\Pr,$ $\complexity_1, \ldots, \complexity_m,
$([m], \M,T, \Pr,$ $\complexity_1, \ldots, \complexity_m,
u_1, \ldots, u_m)$, where 
\beginsmall{itemize}
%joe21
%\item $[m]$ is the set of players;
\item $[m] = \{1, \ldots, m\}$ is the set of players;
%\item each player $i$'s type space is $\{0,1\}^m$;
%joe8*: do we want to restrict the type space to a finite subset of
%{0,1}^*, to make this a finite game?
%joe21: don't need this; covered by the type statement below
%\item each player $i$'s and nature's type space is $\{0,1\}^*$;
\item $\M$ is the set of possible
%joe10*
machines;
%raf30: do we need closed under comp?
%joe13*: we do if you ever define one machine as a composition of two
%others, which I think we did at one point.  If we don't any more, it's OK
%machines, which we assume is closed under composition;
%\item for each $i\in [m]$, a finite set $A_i$;
%(the set of \emph{actions} for player $i$);
%joe6: 
%\item $\Pr$ is a distribution on $\T = \{0,1\}^{m(n+1)}$, the set of type
%\item $\Pr$ is a distribution on $\T = \{0,1\}^{mn}$, the set of type
%joe8*: reinstated the next line,  which I suspect was inadvertantly
%commented out.  Do we want to insist the \Pr has finite support?

%raf32: added T
%joe15: removed comma
%\item $\T \subseteq (\{0,1\}^{*})^{m+1}$, is the set of type 
\item $\T \subseteq (\{0,1\}^{*})^{m+1}$ is the set of type 
%joe13
%profiles, where the $m+1$'th element in the profile corresponds to
profiles, where the $(m+1)$st element in the profile corresponds to
nature's type;%
%\fullv{
%raf94: remove joe79*
%\footnote{We may want to restrict the type space to be a finite subset
%of $\{0,1\}^*$ (or to restrict $\Pr$ to having finite support, as is
%often done in the game-theory literature, so
%that the type space is effectively finite) or to restrict the output
%space to being finite, although these assumptions are not 
%essential for any of our results.}
%}

%raf32: changed to T, made this changed globally
%\item $\Pr$ is a distribution on $\T = (\{0,1\}^{*})^{m+1}$, the set of type 
\item $\Pr$ is a distribution on $\T$;
%joe2
%\item $\size_i: \M_i \rightarrow \IN$ describe the size (or
%joe4
%\item $\size: \union_{i=1}^n \M_i \rightarrow \IN$ describe the size (or
%complexity) of each machines;
%\item $\timec: (\union_{i=1}^n \M_i) \times \{0,1\}^m \times
%\{0,1\}^\infty
%joe5*: here I'd prefer to separate the type and random string.  Is
%there an advantage in not doing so?
%\item $\complexity_i: \M \times  \times\{0,1\}^\infty) 
%\rightarrow \IN$ describes the complexity of a given machine in a given view. 
%joe6: OK, back to views
%\item $\complexity_i: \M \times \{0,1\}^* \times \{0,1\}^\infty
%joe7: back again; views are now finite
%\item $\complexity_i: \M \times \{0,1\}^\infty
%raf36
%\item $\complexity_i: \M \times \{0,1\}^*
\item $\complexity_i$ is a complexity function;
%joe10*: OK; let's make it a real
%\rightarrow \IN$ describes the complexity of a given machine given a
%joe15*: Rafael, do we want the complexity to be real or natural?  It seems to
%me that later we assume it's natural.
%raf33: i prefer natural
%joe16*: OK; I'll change it everywhere
%\rightarrow \IR$ describes the complexity of a given machine given a
%raf36: remove
%\rightarrow \IN$ describes the complexity of a given machine given a
%view; 
%joe2*: here I think we need the cost to depend on the player; 
%\item $\switch_i: \M_i \rightarrow \IN$ describes the cost
%for player $i$ of implementing (or switching to) a given machine;
\item 
%joe4
%$u_i: (\{0,1\}^{mn} \times (\{0,1\}^*)^n \times \IN^n \times \IN^n
%\times
%joe6: taking anture's type into account
%$u_i: (\{0,1\}^{mn} \times (\{0,1\}^*)^n \times \IN^n) 
%joe10*: again; note that we may have an issue when we have to deal with
%computable utilities
%$u_i: (\{0,1\}^*)^n \times (\{0,1\}^*)^n \times \IN^n) 
%joe13*: first, changed * to \infty to allow from infinite computation.
%Should we say something about whether utilities are computable?
%$u_i: (\{0,1\}^*)^{m+1} \times (\{0,1\}^*)^m \times \IR^m) 
%raf32: changed to T
%$u_i: (\{0,1\}^*)^{m+1} \times (\{0,1\}^\infty)^m \times \IR^m) 
%raf37: changed to IN
%$u_i: \T \times (\{0,1\}^\infty)^m \times \IR^m)
%raf42: 
%$u_i: \T \times (\{0,1\}^\infty)^m \times \IN^m) 
%raf47
%$u_i: \T \times (\{0,1\}^\infty)^m \times \IN^m
$u_i: \T \times (\{0,1\}^*)^m \times \IN^m
\rightarrow \IR$ is player $i$'s utility function.   
%joe4
%Intuitively, $u_i(\vec{t},\vec{a},\vec{c}, \vec{d}, e)$ is the
Intuitively, $u_i(\vec{t},\vec{a},\vec{c})$ is the
utility of player $i$ if $\vec{t}$ is the type profile, $\vec{a}$ is the
action profile (where we identify $i$'s action with $M_i$'s output), and
%joe4
%$vec{c}$ is the profile of machine sizes, $\vec{d}$
%is the profile of machine running times, and $e$ is the
$\vec{c}$ is the profile of machine complexities.  
\endsmall{itemize}
\ED
%Just as we can talk about $i$'s expected utility if a given strategy
%profile is played in a  Bayesian games, in a Bayesian machine game, we
%can talk about $i$'s expected utility given a tuple of machines in a
%straightforward way.  
%joe5: done
%\Rnote{We should note that the definition of a game is ``redundant'' in
%that $\complex$ 
%is defined for all input length.} 
%joe6: rewrote
%Note that the, although we have restricted the type space to
%$\{0,1\}^m$, the output of the machine is defined for types of all
%lengths; similarly, the complexity function and utility function are
%defined for types of all lengths.  
%raf
%Like machines, the complexity function and utility function are defined
%for all possible types, not just types in $\{0,1\}^m$ (although all that
%matters in a particular game is the way these function works for the
%type space in that game).  
%When computing the utility in a game
%with type space $\{0,1\}^m$, all that matters in the behavior of
%machines and these functions on types in $\{0,1\}^m$.  
%Defining these
%functions more generally is in part a technical device that makes it
%easier for us to compare different games.  However, this also seems to
%us conceptually appropriate to view a machine as a device that can act
%no matter what its input, and for its complexity and utility to be
%defined in terms of how it acts in general. 
%raf30: added type for nature.

We can now define the expected utility of a machine profile.
%Given a Bayesian machine game $G = ([m], m,
Given a Bayesian machine game $G = ([m],
%joe4
%\M,\Pr, \timec, \size, \switch, \vec{u})$, $\vec{t} \in \{0,1\}^{mn}$,
%\M,\Pr, \vec{\complexity}, \vec{u})$, $\vec{t} \in \{0,1\}^{mn}$,
%raf32 added below, did not always mark changes
%raf94: joe79
%\M,\Pr, T, \vec{\complexity}, \vec{u}), \vec{t} \in \T$,
\M,\Pr, T, \vec{\complexity}, \vec{u})$
%joe21
%and $\vec{M} \in \M^m$, define the random variable $u^{\vec{M}}_i$ on 
and $\vec{M} \in \M^m$, define the random variable $u^{G,\vec{M}}_i$ on 
%joe2
%$\{0,1\}^{mn} \times \{0,1\}^{mn}$ (i.e., the space of type profiles and
%$\{0,1\}^{mn} \times (\{0,1\}^{\infty})^n$ (i.e., the space of type
%$(\{0,1\}^*)^{m+1} \times (\{0,1\}^{\infty})^m$ (i.e., the space of type
$\T \times (\{0,1\}^{\infty})^m$ (i.e., the space of type
profiles and 
sequences of random strings) by taking
%raf86: smaller in shortv
\fullv{
$$
%\begin{array}{ll}
%&u_i^{G,\vec{M}} (\vec{t}, \vec{r}) 
%\\
%= &u_i(\vec{t},
%joe4: put it on a single line
u_i^{G,\vec{M}} (\vec{t}, \vec{r}) = u_i(\vec{t},
%joe4
%(M_j(t_j,r_j))_{j = 1, \ldots, m}, (\size(M_j))_{j= 1, \ldots, m}, 
%(\timec(M_j(t_j,r_j))_{j=1, \ldots, m}, \switch(M_i)).
%joe30: I have a sight preference for writing it this way
%(M_j(t_j;r_j))_{j = 1, \ldots, m}, (\complexity_j(M_j, t_j; r_j))_{j=
%1, \ldots, m}).  
M_1(t_1;r_1), \ldots, M_m(t_m;r_m), \complexity_1(M_1, t_1; r_1),
\ldots, \complexity_m(M_m, t_m)).  
%joe5: a leftover
%\switch(M_i)).
%\end{array}
$$
}
\shortv{$u_i^{G,\vec{M}} (\vec{t}, \vec{r}) = \linebreak u_i(\vec{t},
M_1(t_1;r_1), \ldots, M_m(t_m;r_m), \complexity_1(M_1, t_1; r_1),
\ldots, \complexity_m(M_m, t_m)).$}
%raf47**: complexity should only be defined using the *actual* parts of
%randomness and type read... 
%joe30: dealt with above
%joe17*: a first attempt at notation
Note that there are two sources of uncertainty in computing the expected
utility: the type $t$ and realization of the random coin tosses of the
players, which is an element of $(\bit^\infty)^k$.  
Let $\Pr_U^k$ denote the uniform distribution on $(\{0,1\}^\infty)^k$.
%joe21
%Given an distribution arbitrary distribution $\Pr_X$ on a space $X$, we
Given an arbitrary distribution $\Pr_X$ on a space $X$, we
%raf42:
%write $\Pr_X^{+k}$ to denote the distribution $\Pr \times \Pr_U^k$ 
write $\Pr_X^{+k}$ to denote the distribution $\Pr_X \times \Pr_U^k$ 
on $X \times (\{0,1\}^\infty)^k$.  If $k$ is clear from context or not
relevant, we often omit it, writing $\Pr_U$ and $\Pr_X^+$.
%Given the
%distribution $\Pr$ on the type space $\T$, %= (\{0,1\}^{*})^{n+1}$, 
%let $\Pr^+$
%denote the distribution on $\T \times (\{0,1\}^\infty)^m$ that is the
%product of $\Pr$ and the uniform distribution on tuples of random
%strings.    
Thus, given the probability $\Pr$ on $T$, the expected utility 
of player $i$ in game $G$ if $\vec{M}$ is used is the expectation of the
%joe21
%random variable  $\vec{u}^{G,\vec{M}}_i$
random variable  $u^{G,\vec{M}}_i$
with respect to the distribution $\Pr^+$ (technically, $\Pr^{+m}$):
%joe17
%We can then define the expected utility to player $i$ of
%$\vec{M}$ in game $G$ is
%raf72: same line
$U^G_i(\vec{M}) = \mathbf{E}_{\Pr^+ }[u_i^{G,\vec{M}}].$
%$$U^G_i(\vec{M}) = \mathbf{E}_{\Pr^+ }[u_i^{G,\vec{M}}].$$ 
%that is, as the expectation of the random variable 
%joe10
%$\vec{u}^{\vec{M}}_i$
%$\vec{u}^{G,\vec{M}}_i$
%with respect to the distribution $\Pr^+$.
%joe103
%Note that 
This notion of utility allows an agent to prefer a machine
that runs faster to one that runs slower, even if they give the same
output, or to prefer a machine that has faster running time to one that
gives a better output.
%joe56*: added rest of paragraph 
Because we allow the utility to depend on the whole profile of
complexities, we can capture a situation where $i$ can be ``happy'' as
long as his machine runs faster than $j$'s machine.  Of course, an
important special case is where $i$'s utility depends only on his own
complexity.  All of our 
%raf94:
%technical 
results continue to hold if we make
this restriction.

%raf56:
\subsection{Nash Equilibrium in Machine Games}
%raf56:
%Given this definition of utility, we can now define
%joe37
%Given the above definition of utility, we can now define
Given the definition of utility above, we can now define
($\epsilon$-) Nash equilibrium in the standard way.

\BD [Nash equilibrium in machine games]
%raf73: needed M
%Given a Bayesian machine game $G$, 
Given a Bayesian machine game 
$G = ([m], \M,T, \Pr,$ $\vec{\complexity},\vec{u})$,
%raf73:
%a machine profile $\vec{M}$, 
a machine profile $\vec{M} \in \M^m$,
and $\epsilon \ge 0$, 
$M_i$ is an $\epsilon$-best response to $\vec{M}_{-i}$ 
if, for every $M_i' \in \M$, 
%raf86: shortv
%raf94: joe80 typos
%\fullv{$$U_i^G[(M_i, \vec{M}_{-i}] \geq U_i^G[(M_i', \vec{M}_{-i}] -
\fullv{$$U_i^G[(M_i, \vec{M}_{-i})] \geq U_i^G[(M_i', \vec{M}_{-i})] -
\epsilon.$$} 
%joe80: again, just in case
%\shortv{$U_i^G[(M_i, \vec{M}_{-i}] \geq U_i^G[(M_i', \vec{M}_{-i}] -
\shortv{$U_i^G[(M_i, \vec{M}_{-i})] \geq U_i^G[(M_i', \vec{M}_{-i})] -
- \epsilon.$}
(As usual, $\vec{M}_{-i}$ denotes the tuple consisting
of all machines in $\vec{M}$ other than $M_i$.)
%joe21
%$\vec{M}$ is an \emph{$\epsilon$-Nash equilibrium} for $G$ if, 
$\vec{M}$ is an \emph{$\epsilon$-Nash equilibrium} of $G$ if, 
for all players $i$, $M_i$ is an $\epsilon$-best response to
$\vec{M}_{-i}$. A \emph{Nash equilibrium} is a 0-Nash equilibrium.
\ED
%raf107*: added for AE
Note that in our definition of a best response for player $i$ we
consider the 
%joe104
%ex-ante notion of expected utility, rather than conditioning on
ex ante expected utility, rather than conditioning on
player $i$'s type $t_i$. In standard Bayesian games both notions lead
to an equivalent definition of Nash equilibrium, but as we discuss in
Example \ref{primality.ex}, this is no longer the case for machine
games.

%joe56*
There is an important conceptual point that must be stressed with regard
to this definition.  Because we are implicitly assuming, as is standard
in the game theory literature, that the game is common knowledge, 
we 
%raf107
%are
assume that the agents understand the costs associated with each Turing
machine.  That is, they do not have to do any ``exploration'' to compute
the costs.  In addition, we do not charge the players for
computing which machine is the best one to use in a given setting; we
assume that this too is known.  This model is appropriate in settings
where the players have enough experience to understand the behavior of
all the machines on all relevant inputs, either through experimentation
or theoretical analysis.   We can easily extend the model to incorporate
uncertainty, by allowing the complexity function to depend on the state
%raqf95:
%(type) of nature as well as the machine and the input; see
(type) of nature as well as the machine and the input.
%Example~\ref{primality.ex} for further discussion of this point. 
%raf95: added as per your suggestion below.
For example, if we take complexity to be a function of running time, a
machine $M$ has a true running time for a given input, but a player may
be uncertain as to what it is.  We can model this uncertainty by
allowing the complexity function to depend on the state of nature. 
To simplify the exposition, we do not have the complexity function
%joe92: changed HP10 to HP10a globally
depend on the state of nature in this paper; we do so in \cite{HP10a},
%joe90
%where the uncertainy plays a more central role.  However, we remark that
where the uncertainty plays a more central role.  However, we remark that
it is straightforward to extend all the results in this paper to the
setting where there is uncertainty about the complexity and the
``quality'' of a TM.

%here.  We can say that the complexity is \emph{objective} if it is
%indepedendent of the state, and use objective complexity when
%necessary.  But I don't see the problem in the definition of positive
%state cost; we can just carry along the state everywhere; e.g., require
%that \C_i(M^{-q},s,v) < \C_i(M,s,v) for all s.
%player 1 complexity is defined as \complex_i(t,M_1,v). doesn't this
%mean that player 1 and 2 have the same 
%belief about player 1's complexity?
%We can also capture that different players have different beliefs about the
%complexity of a certain algorithm by allowing the complexity function to
%depend on the identity of the player. For instance, player 1 and 2 might
%have different beliefs about player 3's complexity for running the
%machine $M$ on a given input. 

%raf94*: should we do this or reference the decision theory paper?
%joe85*: See my earlier joe85*.
%joe79*: Rafael, I think we may want to say a little more here, at least
%in the full paper.  Essentially, this gives us a way of modeling
%impossible possible worlds.  
%For example, if I take the complexity to
%be a function of running time, the machine has a true running time, but
%the agent may be uncertain as to what it is.  We model this by allowing
%the running time to depend on the state of nature.  
%\fullv{(type) of nature as well as the machine and the input.}

%raf72: put this back in appendix
\fullv{
%joe5*
%joe56:
%It is worth noting that, in this framework, an $\epsilon$-Nash
One immediate advantage of taking computation into account is that we
can formalize the intuition that $\epsilon$-Nash equilibria are
reasonable, because players will not bother changing strategies for a gain
of $\epsilon$.  Intuitively, the complexity function can ``charge''
$\epsilon$ for switching strategies.  Specifically,
an $\epsilon$-Nash
equilibrium $\vec{M}$ can be converted to a Nash equilibrium by 
%joe56
%slightly modifying 
%%joe10
%%the $\complexity$ function and utility function.  We have player $i$'s
%the complexity and utility functions.  We simply have player 
%$i$'s complexity function incorporate the overhead of switching from
modifying player $i$'s complexity function to incorporate the overhead
of switching from 
$M_i$ to some 
other strategy, and having player $i$'s utility function decrease by
$\epsilon' > \epsilon$ if the switching cost is nonzero; we omit the
formal details here.  Thus, the framework lets us incorporate explicitly
the reasons that players might be willing to play an $\epsilon$-Nash
equilibrium.
%raf72: added this in the appendix (where this section is), also added
%here for consitency 
%joe66
%The above justification of $\epsilon$-Nash equilibrium
This justification of $\epsilon$-Nash equilibrium
seems particularly appealing when designing mechanisms (e.g., cryptographic protocols) where the
equilibrium strategy is made ``freely'' available to the players (e.g., it is accessible on a web-page), but any other strategy requires some 
implementation cost.  
}%fullv
%joe2*: new, based on yesterday's discussion

%raf32: added, and rewrote
Although the notion of Nash equilibrium in Bayesian machine games is
defined in the same way as Nash equilibrium in standard Bayesian games,
%joe15
%the introduction of complexity a leads to some significant differences in
the introduction of complexity leads to some significant differences in
their properties. 
%raf43:
We highlight 
%raf53: 
a few 
%three 
of them here.  
First, note that our definition of a Nash equilibrium considers 
%joe25
%so called
\emph{behavioral strategies}, which in particular might be
randomized. 
%joe25: it's not an alternative
%An alternative definition of Nash equilibria, instead
%considers \emph{mixed strategies}, which are probability
%distributions of deterministic strategies. In traditional game theory
It is somewhat more common in the game-theory literature to consider
\emph{mixed strategies}, which are probability
distributions over deterministic strategies.   As long as agents have
perfect recall, 
%joe25: not quite
%both definitions are equivalent. 
mixed strategies and behavioral strategies are essentially equivalent
\cite{Kuhn53}. 
%raf44: for bayesian games?
%joe26: I believe that Kuhn was considered arbitrary extensive-form games. 
%Since he would allow an initial move by nature, this subsumes Bayesian
%games. 
%raf45: i just wondering about the ``essentially'' requirement for the
%case of Bayesian games. 
%joe26: The essentially is because there is an issue as to what
%``equivalent'' means
However, in our setting, since we might want to charge  
for the randomness used in a computation, such an equivalence 
%joe25
%might no longer hold. 
does not necessarily hold.

%joe25*
%%raf43: put back from Appendix
%Secondly,
Mixed strategies are needed to show that Nash equilibrium always exists
in standard Bayesian games.
%The first is that, 
As the following example shows,
%joe25
%unlike Bayesian games, 
since we can charge for randomization,
a Nash equilibrium may not exist in 
%joe25
%Bayesian machine games, 
a Bayesian machine game,
%joe13*: Rafael, there's a minor subtlety here.  A Nash equilibrium may
%not exist either if the space of actions isn't compact (and I don't
%think ours is in any natural topology), 
%joe66: line shaving
%even if we restrict our attention to games where the type space and the
even in games where the type space and the
output space
are finite.
%joe13
%joe25: now unnecessary
%and we allow randomization.
%joe13*
%\footnote{This follows from the fact that if randomization
%is expensive we end up in a situation where only pure (i.e.,
%deterministic) strategies 
%can be used. Furthermore it is well known that there exists games with
%no pure strategy Nash Equilibrium.} See Appendix for more details. 
%\Rnote{put back the example?}
%joe13* reinstated for now; it's a pretty cute example.  We can
%certainly cut it in the abstract
%\fullv{
%as the following example shows.
%raf72:
%\begin{example}
\begin{example} [Rock-paper-scissors]
\label{roshambo}
 {\rm
Consider the 2-player Bayesian game of roshambo (rock-paper-scissors).
Here the type space has size 1 (the players have no private information).  
We model playing rock, paper, and scissors as playing 0, 1, and 2,
respectively.  The payoff to player 1 of the outcome $(i, j)$ is 1 if $i
= j \oplus 1$ (where $\oplus$ denotes addition mod 3), $-1$ if $j = i
\oplus 1$, and 0 if $i = j$.  Player 2's playoffs are the negative of those
of player 1; the game is a zero-sum game.  As is well known, the unique
Nash equilibrium of this game has the players randomizing uniformly
between 0, 1, and 2.  }

%raf34, removed ``first''.
%{\rm Now consider a machine game version of roshambo.  First suppose
%that we take  
{\rm Now consider a machine game version of roshambo.  Suppose that we 
%raf43: changed from 0->1
%take the complexity of a deterministic strategy to be 0, and the
%complexity of a 
take the complexity of a deterministic strategy to be 1, and the
complexity of 
%joe35: typo
%a 
%joe4
%the size of a determinstic strategy to be 0, and the size of a
%the complexity of a deterministic strategy to be 0, and the complexity of a
%nondeterministic strategy to be 1, and take player $i$'s utility to be
%raf43: nondet -> randomzing
%joe25
%randomizing strategy to be 2, and take player $i$'s utility to be
%joe51
%joe103: reinstated
a 
strategy that uses randomization to be 2, and take player $i$'s
utility to be 
%joe12*
%the sum of his payoff in the underlying Bayesian game and the complexity
his payoff in the underlying Bayesian game minus the complexity
of his 
strategy. 
%joe4
%(Thus, we are not charge for computational steps or for implementing the
%strategy.)  
Intuitively, programs involving randomization are more
complicated than those that do not randomize.  With this utility
function, it is easy to see that there is no Nash equilibrium.  For
suppose that $(M_1, M_2)$ is an equilibrium.  
%joe51
%If $M_1$ is randomized, 
If $M_1$ uses randomization, 
then 1 can do better by playing the deterministic strategy $j\oplus 1$,
where $j$ is the action that gets the highest probability according to
%joe51
%$M_2$ (or is the deterministic choice of player 2 if $M_2$ is not
%randomized).  Similarly, $M_2$ cannot be randomized.  But it is well
$M_2$ (or is the deterministic choice of player 2 if $M_2$ does not use
randomization).  Similarly, $M_2$ cannot use randomization.  But it is well
known (and easy to check) that there is no equilibrium for roshambo with 
deterministic strategies. 
%joe13*: cut remainder of the example; it feels like overkill for this paper
%raf35: maybe we want to put it in the appendix; it is kind of cute and
%interesting. also, when i tell people the first example, they directly
%start asking about epsilon-nash. maybe we should mention that in the
%latter example there is a natural eps nash, but the question is whether
%people ``play'' it, as there are world championships. 
%joe19: it's fine with me to put it in the appendix
%raf36: put it back for now we can move to apendix later.
%\commentout{
%joe21: removed paragraph break and made a parenthetical comment
(Of course, there is nothing special about the 
%raf43
%costs of 0 and 1 for deterministic vs.~randomized strategies.  This
costs of 1 and 2 for deterministic vs.~randomized strategies.  This
argument works as long as all three deterministic strategies have the
same cost, and it is less than that of a randomized strategy.)
%joe85
}

%joe37
%Now consider the variant where we do not charge for the
{\rm 
%raf94; joe79: added a sentence 
%joe108
%The non-existence of Nash euqilibrium does not depend on charging
The non-existence of Nash equilibrium does not depend on charging
directly for randomization; it suffices that it be difficult to compute
the best randomized response.
%joe103
%Now 
Consider the variant where we do not charge for the
first, say, 10,000 steps of computation, and after that 
%joe12
%the cost of computation increases.  
there is a positive cost of computation.
It is not hard to show that, in finite
time, using coin tossing with equal likelihood of heads and tails, we
cannot exactly compute a uniform distribution over the
three choices, although we can approximate it 
%raf50:added footnote
closely.\footnote{Consider a probabilistic Turing machine $M$ with
%joe33
%running-time bounded by $T$ which outputs $0$ (resp $1,2$) with
%probability $1/3$. As $M$'s running-time 
%raf94:
%joe79: simplified argument
%running time bounded by $T$ that outputs $0$ (resp $1,2$) with
%probability $1/3$. Since $M$'s running time 
running time bounded by $T$ that outputs either $0$, $1$, or $2$.
Since $M$'s running time 
is bounded by $T$, $M$ can use at most $T$ of its random bits.
%joe66
%there
%thus exists some natural number $p_0$ such that $M$ outputs $0$ for
%$p_0$ out of the $2^T$ possible random strings it receives as
%input. But, since $2^T$ is not divisible by 3, this is a
%contradiction.} 
%joe79
%But if $M$ outputs 0 for $m$ of the $2^T$ strings of length $T$, then its
If $M$ outputs 0 for $m$ of the $2^T$ strings of length $T$, then its
probability of outputting 0 is $m/2^T$, and cannot be $1/3$.}
%joe66: removed paragraph break
%raf94: removing old
\iffalse
running time bounded by $T$ that outputs $0$ (resp $1,2$) with
probability $1/3$. Since $M$'s running time 
is bounded by $T$, $M$ can use at most $T$ of its random bits; there thus exists some natural number $p_0$ such that $M$ outputs $0$ for $p_0$ out of the $2^T$ possible random strings it receives as input. But, since $2^T$ is not divisible by 3, this is a contradiction.}
%joe5*: added
%raf94:fi
\fi
%joe103
%(For example, if we toss a coin twice, playing rock if we get heads
For example, if we toss a coin twice, playing rock if we get heads
%joe85: it's not such a small probability
%twice, paper with heads and tails, scissors with tails and heads, and
%try again with two tails, we do get a uniform distribution, except for
%the small probability of nontermination.)
twice, paper with heads and tails, and scissors with tails and heads, and
try again with two tails, we do get a uniform distribution, conditional
%joe90
%on terminatation.
on termination.
%\Jnote{Rafael,
%do you have a reference?  Should we discuss how you can get it exactly
%with an unbounded machine?}  
%raf94: joe79
%From this
Since there is always a deterministic best reply that can be computed
quickly (``play rock'', ``play scissors'', or ``play paper''), 
from this
observation it easily follows, as above, that there is no Nash
equilibrium in this game either.
%joe37
%joe85
%}
%raf50:
%joe33* Rafael, I found this sentence very hard to  parse.  I don't
%think you're saying that in the games below all computation is free;
%just bounded computation (that is, for example, computation bounded by
%some fixed polynomial of the input).  First, if that's what you're
%trying to say, then I don't think it's being said all the clearly.
%Second, I'm not sure why that's interesting enough to make a fuss
%about.  My temptation would be to cut the next sentence.
%raf51: the only reason was that people in the crypto world have posed
%it as an open problem... 
%As a corollary we also get that there exists games without Nash Equilibria 
%joe34*: I still found this hard to parse; when you say ``bounded
%running time'', do you mean bounded by a constant?  
%As a corollary, we also get that there exists games without a Nash
%equilibrium where computation is \emph{free}, as long as we restrict
%our attention 
%joe108
%As a corollary, it follows that that there are computational
As a corollary, it follows that there are computational
games without a Nash
equilibrium where all constant-time strategies are taken to be free.
%our attention 
%to strategies that  
%%joe33
%%can be computable by randomizing Turing machine with bounded
%%running-time.
%can be computed by randomizing Turing machine with bounded running time.
%raf50*: an interesting question would be to show a game without NE even
%when considering expected poly strategies. probably this is know...
%raf52*: since we claim our framework explains real-life observation, we 
%should probably mention championships; do we need to talk about
%repeated games?  
%joe52*: How about adding some sentences like the following somewhere:
%raf53*: great! need to find refs..
It is well known that people have difficulty simulating randomization;
%[we could use a reference for this]; 
we can think of the cost for
%raf53:typo
%randomizing as capturing this difficult.  Interestingly, there are
randomizing as capturing this difficulty.  Interestingly, there are
roshambo tournaments (indeed, even a Rock Paper Scissors World
Championship), and books written on roshambo strategies 
%joe53: I added this to joe.bib
%[there's a book sold at Amazon that we can reference]. 
\cite{Walker04}.
Championship players are clearly
not randomizing uniformly (they could not hope to get a higher payoff
than an opponent by randomizing).  Our framework provides a
psychologically plausible account of this lack of randomization.
%raf36: put back you old question. maybe we should mention tournaments?
%\Jnote{Should we say something about finitely repeated roshambo?  This
%is what they play in tournaments.}
%joe8: added to mark end of example
%joe85
}

%joe21*
%joe85
{\rm
The key point here is that,
in standard Bayesian games, to guarantee equilibrium requires using
randomization.  Here, we allow randomization, but we charge for it.
This charge may prevent there from being an equilibrium.
}
\qed 
\end{example}

%raf94: this is a text from the ICS paper that we might want to put here
%instead of the above para, or just drop. 
%joe85: Text like this should go before the theorem; this seems like the
%wrong place.  But we do have something along those lines in line 2875
\iffalse
In these examples, computing an appropriate randomized response was
expensive, either because we charged directly for randomizing, or
because getting the ``right'' response required a lot of computation.
It turns out that the cost of randomization
is the essential roadblock to getting a Nash equilibrium in Bayesian
machine games. We make this precise in the full version of the paper.
\fi

Example~\ref{roshambo} shows that sometimes there is no Nash
equilibrium.  It is also trivial to show that given any standard
Bayesian game $G$ (without computational costs) and a \emph{computable}
strategy profile  $\vec{\sigma}$ in $G$ (where $\vec{\sigma}$ is
computable if, for each player $i$ and
type $t$ of player $i$, there exists a Turing machine $M$ 
that outputs $a$ with probability $\sigma_i(t)(a)$),
 we can choose computational costs and modify the utility
%joe102
%function in $G$ in such a way as to make $\vec{\sigma}$ an equilibrium
function in $G$ in such a way as to make $\vec{\sigma}$ a
dominant-strategy equilibrium 
of the modified game:  
%raf104*: i am not convinced by this; it is trivial to say
%that any strategy is an eq if we allw to modfy the game; not quite
%sure how to change it.
%joe102: I think the point here is that we can make \sigma_i a dominant
%strategy for player i by making every other strategy costly.  While
%this is a trivial point, it does explain hysteresis effects.
%Not only is there sometimes no Nash equilibrium, given any strategy
%profile $\vec{\sigma}$ in a Bayesian game where, for each player $i$ and
%type $t$ of player $i$, $\sigma_i(t)$ induces a computable distribution
%over actions
%(i.e.,  a distribution $\Pi$ for which there exists a machine $M$ 
%that outputs $a$ with probability $\Pi(a)$)
%we can always extend the Bayesian game to a Bayesian machine game 
%joe5
%where the distribution with a Nash equilibrium that has the same distribution
%over actions.  
%with a Nash equilibrium $\vec{M}$ where $M_i$ implements $\sigma_i$
%(i.e., it induces the same distribution over actions for each type).
%This can in fact be done easily: 
we simply make the cost to
player $i$ of implementing a strategy other than $\sigma_i$ sufficiently
high.

One might be tempted to conclude from these examples that Bayesian
machine games are uninteresting.  They are not useful prescriptively,
since they do not always provide a prescription as to the right thing to
do.  Nor are they useful descriptively, since we can always ``explain''
the use of a particular strategy in our framework as the result of a
high cost of switching to another strategy.  With regard to the first
%joe25: we don't really show this
%raf44: our def of security provides prescriptive insight in some sense...
%point, as we will show, our framework can provide useful prescriptive
point, 
%joe26: how about this
%we believe that our framework can provide useful prescriptive
as 
%raf104*:
\iffalse
shown by our definition of security, our framework can provide useful
prescriptive 
%joe12
%insights by making minimal assumptions on the form of the 
insights by making minimal assumptions regarding the form of the 
%joe10
%$\complexity$
complexity
%,$\size$, and $\switch$ 
%joe25
%functions.  
function.
\fi
%raf104
we show in a companion paper \cite{HP09b}, our framework can provide useful
prescriptive insights in the context of cryptographic protocols.
%joe6*: added
Moreover, although there may not always be a Nash equilibrium, 
it is easy to see that 
there is always an $\epsilon$-Nash equilibrium for some $\epsilon$;
this $\epsilon$-Nash can give useful
guidance into how the game should be played.  For example, in the second
variant of 
the roshambo example above, we can get an $\epsilon$-equilibrium
for a small $\epsilon$ by attempting to simulate the uniform
distribution by tossing the coin 10,000 times, and then just playing
rock if 10,000 tails are tossed.  Whether it is worth continuing the
simulation after 10,000 tails depends on the cost of the additional
computation.
%joe56: added
Finally, as we show below, there are natural classes of games where a
Nash equilibrium is guaranteed to exist (see Section~\ref{sec:suff}).
With regard to the second point, we
would argue that, in fact, it is the case that people continue to play
certain strategies that they know are not optimal because of the
overhead of switching to a different strategy; that is, our model
captures a real-world phenomenon.  
%raf50*: additionally, maybe use ``simple'' description/occam's razor to
%be descriptive ...e.g., roshambo/repeated prisoners
%dilemma...primes. we mention the primes example later but it nicely
%shows that our framework is descriptive. it would be even nicer if we
%could give a simple version of the repeated prisoners dilemma with weak
%assumptions about cost of comp. for instance, if we view a strategy as
%a TM that at each stages receives as input the previous move of the
%opponent and outputs its current move, and let stateless machines be
%cheaper than stateful ones, then ``tit-for-tat'' will be a Nash Eq in
%the rep prisoners dilemma (with appropriate discounting). was this
%already what rubinstien showed? 
%}%commentout
%joe5: cut this for now
%\Rnote{Maybe add something about the fact tha $\epsilon$ equilibria
%exist, and that 
%a ``good'' solution concept (at least in the ``prescriptive'' sense) 
%would be the $\epsilon$-equilibria with the smallest $\epsilon$/}

%joe13*: added example
%joe21
%raf43:
%A second property of (standard) Bayesian games that does not hold for
%raf72: another
%A third property of (standard) Bayesian games that does not hold for
Another property of (standard) Bayesian games that does not hold for
Bayesian machine games is the following.
%joe21
%a Nash equilibrium $\vec{\sigma}$ in a Bayesian game has the property
%that,
Given a Nash equilibrium $\vec{\sigma}$ in a Bayesian game,
not only is $\sigma_i$ 
a best response to $\vec{\sigma}_{-i}$, 
but it continues to be a best response conditional on $i$'s type.
That is, if $\Pr$ is the probability distribution
on types, and $\Pr(t_i) > 0$,
%raf42
%then $U_i(\sigma_i,\vec{\sigma}_{-i} \mid t_i) \ge U_i(\sigma',
then $U_i(\sigma_i,\vec{\sigma}_{-i} \mid t_i) \ge U_i(\sigma'_i,
%raf56:
%\vec{\sigma}_{-i})$ for all strategies $\sigma'_i$ for player $i$, where 
%raf107*: added a formal def above
%\vec{\sigma}_{-i} \mid t_i)$ for all strategies $\sigma'_i$ for
%player $i$, where 
\vec{\sigma}_{-i} \mid t_i)$ for all strategies $\sigma'_i$ for player
$i$.
%, where 
%$U_i(\sigma_i,\vec{\sigma}_{-i} \mid t_i)$ is the expected utility of 
%$\vec{\sigma}$ conditional on player $i$ having type $i$. 
Intuitively,
if player $i$ could do better than playing $\sigma_i$ if his type were
$t_i$ by playing $\sigma'_i$, then $\sigma_i$ would not be a best
response to $\vec{\sigma}_{-i}$; 
we could just modify $\sigma_i$ to
agree with $\sigma_i'$ when $i$'s type is $t_i$ to get a strategy that
does better against $\vec{\sigma}_{-i}$ than $\sigma_i$.
This is no longer true with Bayesian machine games, as the following
simple example shows.

\begin{example}
%raf72
[Primality guessing]
\label{primality.ex}
 {\rm Suppose that the probability on the type space assigns
uniform 
%joe21
%probability to all $2^{100}$ bit strings of the form $1{0,1}^{100}1$
%(i.e., the odd numbers between $2^{100}$ and $2^{101}$).
probability to all $2^{100}$ odd numbers between $2^{100}$ and $2^{101}$
(represented as bit strings of length 100).
Suppose that a player $i$ wants to compute if its input (i.e.,
its type) is prime.  Specifically, $i$ gets a utility of 2  minus the
costs of its 
running time (explained below) if $t$ is prime and it outputs 1 or if $t$
is composite and it outputs 0; on the other hand, if it outputs either 0
or 1 and gets the wrong answer, then it gets a utility of $-1000$.
But $i$ also has the option of ``playing safe'';
if $i$ outputs 2, then $i$ gets a utility of 1 no matter what the input
is.  The 
running-time cost is taken to be 0 if $i$'s machine takes less than 2 units
of time and otherwise is $-2$.  We assume that outputting a constant
function takes 1 unit of time.  Note that although testing for primality
is in polynomial time, it will take more than 2 units of time on all 
inputs that have positive probability.  Since $i$'s utility is
independent of what other players do, $i$'s best response is to always
output 2.  However, if $t$ is actually a prime, $i$'s best response
conditional on $t$ is to output 1; similarly, if $t$ is not a prime,
$i$'s best response conditional on $t$ is to output 0.  The key point is
that the
machine that outputs $i$'s best response conditional on a type does not
do any computation; it just outputs the appropriate value.}

%joe56*: added next two paragraphs
{\rm Note that here we are strongly using the assumption that $i$
understands the utility of outputting 0 or 1 conditional on type $t$.  
%raf58:typo
%This amounts to saying that if he is playing the game game conditional
This amounts to saying that if he is playing the game conditional
on $t$, then he has enough experience with the game to know whether $t$
%raf58:typo
%is prime.  If we wanted to capture a more general setting where player
is prime.  If we wanted to capture a more general setting where the player
did not understand the game, even after learning $t$, then we could do
this by considering two types (states) of nature, $s_0$ and $s_1$, where,
intuitively, $t$ is composite if the state (nature's type) is $s_0$ and
prime if it is $s_1$.  Of course, 
$t$ is either prime or it is not.  We can avoid this problem by simply
having the utility of outputting 1 in $s_0$ or 0 in $s_1$ being $-2$
(because, intuitively, in state $s_0$, $t$ is composite and in $s_1$ it
is prime) and the utility of outputting 0 in $s_0$ or 1 in $s_1$ being
2.  The relative probability of $s_0$ and $s_1$ would reflect the
player $i$'s prior probability that $t$ is prime.}

{\rm In this case, there was no uncertainty about the complexity; there
was simply uncertainty about whether the type satisfied a certain
property.  As we have seen, we can already model the latter type of
uncertainty in our framework.  To model uncertainty about complexity, we
simply allow the complexity function to depend on nature's type, as well
as the machine and the input.  We leave the straightforward details to
the reader.}
%raf36:added \qed here too
\qed 
\end{example}

%raf53*: added
%joe53: Rafael, what's the Aumann85 reference?
A common criticism (see e.g., \cite{Aumann85}) of Nash equilibria 
%joe53*: Rafael, this is actually not the standard definition of strict
%equilibrium 
%which userandomized strategies is that such equilibria cannot be
%\emph{strict} (i.e., each player's equilibrium strategy is the unique
that use randomized strategies is that such equilibria cannot be
\emph{strict} (i.e., it cannot be the case that each player's
equilibrium strategy gives a strictly better payoff than any other
strategy, given the other players' strategies).
This follows since any pure strategy in the support of the randomized
%joe53
%strategy will also be a best-response. As the example below shows, this
strategy must give the same payoff as the randomized strategy. As the
example below shows, this 
%joe53
%might no longer be the case when considering games with computational
is no longer the case when considering games with computational
costs. 

\begin{example}
%raf69
[Strict Nash equilibrium]
%joe53
%{\rm Consider the same game as in example \ref{primality.ex} except
%that all machines with running-time smaller than $T$ have a
%running-time cost of 0 (and all others $-2$). It might very well be the
%case that, for some 
{\rm Consider the same game as in Example~\ref{primality.ex}, except that
all machines with running time less than or equal to $T$ have a
cost of 0, and machines that take time  greater than $T$ have a cost of
$-2$. It might very well be the case that, for some 
%joe53
%values of $T$, probabilistic primality testing algorithms can be used to
%always determine wheter $t$ is prime or composite, whereas all
%deterministic algorithms take too much time. (Recall that although
values of $T$, there might be a probabilistic primality testing
algorithms that runs in time $T$ and determines with high probability 
%joe103
%determine 
whether a given input $x$ is prime or composite, whereas all
deterministic algorithms take too much time. 
%joe53: should we add a reference to Rabin or Solovay-Strassen?
%(Recall that although
%primality can be tested in deterministic polynomial-time, known
%randomized algorithms are still significantly faster).} 
(Indeed, although deterministic polynomial-time algorithms for primality
testing are known \cite{AKS02}, in practice, randomized algorithms are
used because they run significantly faster.)}
%raf95
\qed
\end{example}

%raf98: moved back this from the appendix
%raf100
%\section{Machine Games with Mediators}\label{sec:mediator}
%joe93
%\section{Machine Games with Mediators and Extensive-form
%raf102: should we change back the name to just Mach games with Med?
%joe96: I think so (and I did that)
\section{Machine Games with Mediators}\label{sec:mediator} 
Up to now we have assumed that the only input a machine receives 
is the initial type.  This is appropriate in a
normal-form game, but does not allow us to consider
%raf107:
%game where
game where
players can communicate with each other and (possibly) with a trusted
mediator. 
We now extend Bayesian machine games 
to allow for communication.
For ease of exposition, we assume that all communication passes between
the players and a trusted mediator.  Communication between the players
is modeled by having a trusted mediator who passes along messages
received from the players.
Thus, we think of the players as having reliable 
communication channels to and from a mediator; no other
communication channels are assumed to exist.

The formal definition of a Bayesian machine game with a mediator is
similar in spirit to that of a Bayesian machine game, but now we assume
that the machines are \emph{interactive} Turing machines, that can
also send and receive messages.  We omit the formal definition of an
interactive Turing machine (see, for example, \cite{goldreich01}); roughly
speaking, the machines use a special tape where the message to be sent
is placed and another tape where a message to be received is written.
The mediator is modeled by an interactive Turing  machine that we denote
$\F$.

A \emph{Bayesian machine game with a mediator} (or a mediated Bayesian
machine game) is thus a pair $(G, \F)$, where $G= ([m],\M,\Pr,
%joe99
%\complexity_1, \ldots, \complexity_n, 
%u_1, \ldots, u_n)$ is a Bayesian machine game (except that $\M$ here denotes 
\complexity_1, \ldots, \complexity_m, 
u_1, \ldots, u_m)$ is a Bayesian machine game (except that $\M$ here denotes 
a set of \emph{interactive} machines) and $\F$ is an interactive Turing
machine. 

Like machines in Bayesian machine games, interactive machines in a game
with a mediator take as argument a view and produce an outcome.  Since 
what an interactive machine does can depend on the
history of messages sent by the mediator, the message history (or, more
precisely, that part of the message history actually read by the
machine) is also part of the view.  
Thus, we now define a view to be a string $t;h;r$ in $\bit^*;\bit^*;\bit^*$,
where, as before,
$t$ is that part of the type actually read and $r$ is a finite bitstring
representing the string of random bits actually used, and $h$ is a finite
sequence of messages received and read.  
Again, if $v = t;\barw;r$, we take $M(v)$ to be the output of $M$
given the view.

We assume that the system proceeds in synchronous stages; a message sent
by one machine to another in stage $k$ is received by the start of stage
$k+1$.  
%joe104
%More formally, following \cite{ADGH06},
More formally, as in the work of Abraham et al.~\citeyear{ADGH06},
we assume that a \emph{stage}
consists of three phases.  
In the first phase of a stage,
each player $i$ sends a message to the
mediator, or, more precisely, player $i$'s machine $M_i$ computes a message
to send to the mediator; machine $M_i$ can also send an
empty message, denoted $\lambda$.  In the second phase, 
the mediator receives the message and 
mediator's machine sends each
player $i$ a message in response (again, the mediator can send an empty
message). 
In the third phase, each player $i$ 
performs an action other than that of sending a message (again, it may
do nothing).
The messages sent and the actions taken can depend on the machine's
message history (as well as its initial type).

We can now define the expected utility of a profile of interactive
machines in a Bayesian machine game with a mediator.   The definition is
similar in spirit to the definition in Bayesian machine games, except
that we must take into account the dependence of a player's actions on
the message sent by the mediator.  
Let $\view_i(\vec{M}, \F, \vec{t}, \vec{r})$ denote the string
%joe103
%$(t_i;\barw_i;r_i)$ 
$(t_i;\barw_i;r_i)$,
where $\barw_i$ denotes the messages received by player $i$ if the machine profile is $\vec{M}$, the
mediator uses machine $\F$, the type profile is $\vec{t}$, and
$\vec{r}$ is the profile of random strings used by the players and the
mediator.  
Given a mediated Bayesian machine game $G'=(G,\F)$,
we can define the random variable $u_i^{G',\vec{M}}(\vec{t},\vec{r})$ as
before, except that now $\vec{r}$ must include a random string for the
mediator, and to compute the outcome and the complexity function, $M_j$
gets as an argument $\view_j(\vec{M}, \F, \vec{t}, \vec{r})$, 
since this is the view that machine $M_j$ gets in this setting.  
Finally, we define 
$U_i^{G'}(\vec{M}) = \mathbf{E}_{\Pr^+}[u_i^{G',\vec{M}}]$ as before,
except that now $\Pr^+$ is a distribution on $\T \times
%joe92
%({\{0,1\}^\infty})^{n+1}$ rather than $\T \times
%({\{0,1\}^\infty})^{n}$, since we must include a random string for the
({\{0,1\}^\infty})^{m+1}$ rather than $\T \times
({\{0,1\}^\infty})^{m}$, since we must include a random string for the
mediator as well as the players' machines.

We can define Nash equilibrium 
%raf100: not needed in this paper
%and computationally robust Nash equilibrium in games with mediators 
as in
Bayesian machine games;
we leave the details to the reader.

%raf98: moved back this example to this section.
As the following example shows, the \emph{revelation principle} no longer holds
once we take computation into account.
%raf107*: added as suggested by the AE (although we do say this very
%early on in the example...
%joe104
%The idea is simple: if direct revelation strategies are "too" costly,
The idea is simple: if direct revelation strategies are ``too costly'', then
the revelation principle no longer holds.
%\paragraph{The revelation principle}
\begin{example}
%raf69:
%joe62
%[Failure of Revelation principle]
[Failure of revelation principle]
\emph{
%raf68:de-emph
%The \emph{revelation principle} is one of the fundamental principles in
The revelation principle is one of the fundamental principles in
traditional implementation theory. A specific instance of it
%joe55: Rafael, what's the Myerson79 reference?  There's a Myerson86 in
%joe.bib that's relevant
%joe103*: Rafael, do we really want to reference Forges here?  I don't
%think of her as the right person for the revelation principle
%joe105
%\cite{Myerson79,F86} stipulates that for every Nash equilibrium in a
\cite{F86,Myerson79} stipulates that for every Nash equilibrium in a
mediated games $(\G,\F)$, there exists a different mediator $\F'$ such
%games $\G$ where players can communicate with a mediator $\F$, there exists a different mediator $\F'$ such
that it is a Nash equilibrium for the players to \emph{truthfully}
report their type to the mediator and then perform the action suggested
by the mediator. 
%raf90:
%raf98: remove
%(A formal definition of mediated games can be found in Appendix \ref{sec:mediator}.)
As we demonstrate, this principle no longer holds when we take
computation into account 
%joe60
(a similar point is made by Conitzer and
%raf63:
%Sandholm \citeyear{CS??}, although they do not use our formal model).
Sandholm \citeyear{CS04}, although they do not use our formal model).
The intuition is simple: truthfully reporting 
%joe55: typos
%your type will not be an equilibrium it is too ``expensive'' send the
your type will not be an equilibrium if it is too ``expensive'' to send the
whole type to the mediator.   
For a naive counter example, consider a game where a player get utility
1 whenever its complexity is 0 (i.e., the player uses the strategy
%joe103: typo, I presume!
%$\bot$) and positive utility otherwise. Clearly, in this game it can
$\bot$) and utility 0 otherwise. Clearly, in this game it can
never be an equilibrium to truthfully report your type to any
mediator. This example is degenerate as the players actually never use
%joe55
%any mediator. In the following example, we consider a game with a Nash
the mediator. In the following example, we consider a game with a Nash
equilibrium where the players use the mediator. }

%raf68:
\emph{
%raf55: took from your email and rewrote a touch
%joe55: a little more rewriting
%Suppose that the type space
%is such that the players types are $n$-bit
Consider a 2-player game where each player's type is an $n$-bit number.
The type space consists of all pairs of $n$-bit numbers 
that either are the same, or that differ in all but at most $k$
%joe55
%places, where $k << n$.  The player receive a utility of 1 if they can
%raf104:
%places, where $k \ll n$.  The player receive a utility of 1 if they can
places, where $k \ll n$.  The players receive a utility of 1 if they can
guess correctly whether their types are the same or not, while having
%raf104:
%communicated less than $k+2$ bits; otherwise it receives a utility of
communicated less than $k+2$ bits; otherwise, they receive a utility of
0. 
%joe103
%Consider a mediator that upon receiving $k+1$ bits from 
%%raf104
%each of 
%the players
%answers back to both players whether the bits received are identical or
Consider a mediator that, upon receiving $k+1$ bits from 
each of the players,
tells the players whether the bit sequences are identical.
With such a mediator it is an equilibrium for the players to provide the
first $k+1$ bits of their input and then output whatever the mediator
tells them. 
However, providing the full type is always too expensive (and can thus
never be a Nash equilibrium), no matter what mediator the players have
access to. }
%raf95
\qed
\end{example}

%joe87: it seems strange to have a one-paragraph subsection as the only
%subsection 
%\subsection{Repeated and Extensive Games}
%raf98*: hi joe, i know we said that we would redefine games with
%mediator to no longer be cheap, and then have cheap mediated games as a
%special case. but, i now feel again that the name ``mediated games''
%should be reserved for games where the messages sent to the mediator
%are cheap (i.e., utility does not depend on the messages, but only the
%final actions). what we are defining here really is an extensive-form
%game.   
%joe87
%We can extend the treatment above to repeated games (where
Repeated games and, more 
generally, arbitrary extensive-form games (i.e., games defined by game
%raf98: 
%trees). We capture an extensive-form game by simply viewing nature as a
%joe87
%trees). Wecapture an extensive-form machine game $G'=(G,\cN)$ by simply
trees) can be viewed as special cases of games with mediators, except
that now we allow the utility to depend on the messages sent to the
mediator (since we
view these as encoding the actions taken in the game).
In more
detail,  we capture an extensive-form machine game $G'=(G,\cN)$ by simply
%raf104: this was commented out
viewing nature as a 
%raf98: 
%mediator;  we allow  utility functions to 
mediator $\cN$;  
we allow 
%raf98
the mediator to be a function of the type of nature, and the 
utility functions to 
%raf98
%take into account the messages sent by the player to the mediator
take into account the messages sent by the players to the mediator
(i.e., talk is not necessarily ``cheap''), and also the random coins
used by the mediator.  
%joe103
%In other words, we assume, without loss of generality, that all
In other words, we assume without loss of generality that all
communication and signals sent to a player are sent through the
mediator, and that all actions taken by a player are sent as messages to
the mediator. 
%raf98: 
%joe87
%Thus, the utility of player $i$, $u_i(\vec{t},h,\vec{c})$ is now a
Thus, the utility of player $i$, $u_i(\vec{t},h,\vec{c})$, is now a
function of the type profile $\vec{t}$, the view $h$ of nature (i.e.,
%joe103
%all the messages received, and random coins tossed, by the mediator),
the messages received and random coins tossed by the mediator),
and the complexity profile $\vec{c}$. 
We leave the details of the formal definition to the
%joe104
%to the 
reader. 

%raf90:
\section{Computational Explanations for Observed Behavior}
\label{sec:behavior}
%raf90: took from shortv
%In particular, 
%joe103
%In this section we show several examples where our framework can
In this section, we give several examples where our framework can
%joe103
%be used to give simple (and, arguably, natural) explanations to
be used to give simple (and, arguably, natural) explanations for
experimentally observed behavior in classical games where traditional
game-theoretic solution concepts fail. 
%joe85
%For instance, we show that in a variant of the \emph{finitely repeated
For instance, we show that in a variant of \emph{finitely repeated
%joe90
%prisonner's dilemma (FRPD)} where players are charged for the use of
prisoner's dilemma (FRPD)} where players are charged for the use of
%joe82: used tit for tat, without hyphens, everywhere
%memory, the strategy \emph{tit-for-tat}---which does exceedingly well in
memory, the strategy \emph{tit for tat}---which does exceedingly well in
experiments~\cite{Axelrod}---is also a Nash equilibrium, whereas it is
dominated by defecting in the classical sense. 
%raf86: wanted to add something like this, but it becomes quite silly
%(there is room for it though) 
%Intuitively, 
%joe76: Why not?  I put it in parens
%raf90
%(This follows from the facts that (1) any profitable
(Intuitively, this follows from the facts that (1) any profitable
deviation from tit for tat requires the use of memory, and (2) if
future utility is discounted, then for sufficiently long games, the
potential gain of deviating becomes smaller than the cost of memory.)
%raf80: this is still very sketchy, i will fix it
%joe74
%Similarly, we show that recently found \emph{first-impression-matters
%joe85: cut, since it's in the other paper
Similarly, 
%we show that \emph{first-impression-matters
%biases} \cite{mrabin98}---i.e., that people tend to put more weight on
%evidence they hear early on---% and 
%joe85: these aren't biases so much as ``imperfections''
%biases observed in real-life randomization (e.g., in sports
% observed in real-life randomization (e.g., in sports competitions) 
%joe90
%the imperfect randomization obesrved in real-life events where
the imperfect randomization observed in real-life events where
randomization helps (e.g., penalty shots in soccer or serving in tennis)
can be explained by considering computational costs;
%joe85: added
as long as a sequence appears random relative to the computation that an
opponent is willing to put in to make predictions about the
sequence, that is good enough.
%raf86: i thought a CS audience might find this interesting 
%joe85: Rafael, what's BM?  Blum-Micali?  Can you send me the reference?
%For instance, in our framework, the computational theory of
%\emph{pseudorandomness} \cite{BM,Yao} can be viewed as having much the
%same motivation as randomization biases.
%joe87*: Rafael, don't forget to give me the BM reference
%raf99: strange that you don't have it (if you have yao), i will resend
Interestingly, the same explanation applies to the computational theory of
\emph{pseudorandomness}.  A sequence is pseudorandom if it passes all
polynomial-time tests for randomness.  There is no need to use a
sequence that is more random than a pseudorandom sequence when playing
against an opponent for whom super-polynomial computation is too
expensive. 
%joe85: moved from below and rewrote slightly
In a companion paper \cite{HP10a}, where we apply our
framework to decision theory, we show how other well-known ``anomalous''
behaviors, such as the \emph{status quo} bias \cite{SZ88} and the 
\emph{first-impressions-matter} bias \cite{mrabin98} can
be explained by viewing computation as a costly resource.
%raf80: need more refs here
%raf95*: these refs are mostly for the bias in information example,
%which we are removing from here? 
%joe85: rewrote slightly
%We mention that the above-mentioned ``anomalies to rationality'' have
%previously been given various cognitive explanations (e.g., using models
Many of these behaviors have been given various cognitive explanations
(e.g., using models of the brain \cite{M98} or using psychology
\cite{mrabin98}).   
%joe85
Some of these explanations can be viewed as being based in part on
considerations of computational cost.
As shown by our results, we do not need to resort to complicated
assumptions about the brain, or psychology, in order to provide such
explanations: extremely simple (and plausible) assumptions about the
cost of computation suffice to explain, at least at a qualitative level,
%joe90
%the observered behavior.  
the observed behavior.  
%joe85
%raf96:
%(See \cite{HP10} and Example~\ref{xam:biases}for further discussion of
(See \cite{HP10a} and Example~\ref{xam:biases} for further discussion of
these points.) 

%raf90: :remove
\iffalse
As shown by the examples below, Nash equilibrium in machine games gives a
plausible explanation of observed behavior in 
%raf88:
well-studied games such as
%joe85
%the finitely-repeated prisoner's dilemma.   
finitely repeated prisoner's dilemma.   
\fi
\begin{example}
%raf69
[Finitely repeated prisoner's dilemma]
\label{xam:pd} 
{\rm Recall that 
in the prisoner's dilemma, there are two prisoners, who
can choose to either cooperate or defect.
As described in the table below, if they both cooperate, they both get
%raf: changed to -1
%3; if they both defect, then both get 1; if one defects and the other
%raf54*:
%3; if they both defect, then both get -1; if one defects and the other
3; if they both defect, then both get -3; if one defects and the other
cooperates, the defector gets 5 and the cooperator gets $-5$.
(Intuitively, the cooperator stays silent, while the defector ``rats
out'' his partner.  If they both rat each other out, they both go to jail.)}
\begin{table}[htb]
\begin{center}
\begin{tabular}{c |  c c}
& $C$ & $D$\\
\hline
$C$ &$(3,3)$ &$(-5,5)$ \\
%joe52
%$D$  &$(5,-5)$  &(-1,-1)  \\
%raf54
%$D$  &$(5,-5)$  &$(-1,-1)$  \\
$D$  &$(5,-5)$  &$(-3,-3)$  \\
\end{tabular}
\end{center}
\end{table}

{\rm It is easy to see that defecting dominates cooperating: no matter what
the other player does, a player is better off defecting than
cooperating.  Thus, ``rational'' players should defect.  And, indeed,
$(D,D)$ is the only Nash equilibrium of this game.  Although $(C,C)$
gives both players a better payoff than $(D,D)$, this is not an
equilibrium.}

{\rm Now consider finitely repeated prisoner's dilemma (FRPD), where
prisoner's dilemma is played for some fixed number $N$ of rounds.  
The only Nash equilibrium is to always defect; this can be seen by a
%joe55: changed step to round throughout
%backwards induction argument.  (The last step is like the one-shot game,
backwards induction argument.  (The last round is like the one-shot game,
so both players should defect; given that they are both defecting at the
last round, they should both defect at the second-last round; and so on.)
This seems quite unreasonable.  And, indeed, in experiments, people do
not always defect \cite{Axelrod}.  In fact, quite often they cooperate
throughout the 
game.  Are they irrational?  It is hard to call this irrational
behavior, given that the ``irrational'' players do much better than
supposedly rational players who always defect.  }

%joe52: added paragraph break
{\rm There have been many
%raf52:
%attempts to explain cooperation in FRPD in the literature (see, for
%example, \cite{KMRW}).  
attempts to explain cooperation in FRPD in the literature; see, for
example, \cite{KMRW,Ney85,PY94}.  
%joe104
%In particular, \cite{Ney85,PY94} demonstrate that if players are
In particular, Neyman \citeyear{Ney85} and Papadimitriou and Yannakakis
\citeyear{PY94} show that if players are 
restricted to 
%joe52
%finite automaton with bounded complexity, then there exists equilibria
%which are different then always defecting; on the negative side, the
%strategies described in those equilibria are quite complex and require
%the equilibria is to force players to remember a short history of the
using a finite automaton with bounded complexity, then there 
%joe54
%exists
exist
equilibria 
that allow for cooperation.  However, the
strategies used in those equilibria are quite complex, and require
the use of large automata;\footnote{The idea behind
%joe86*: I thought that in these equilibra each player had to do a
%raf97**: sure, i will fix this.
%particular (somewhat complicated) sequence of moves, and watched to
%ensure that other players did their moves, punishing if they didn't.
%The idea is that the making the moves and watching the other player
%kept you from counting.  If that's right, I think we should rewrite the
%footnote.  
these equilibria is to force players to remember a short history of the
game, during which players perform random actions; this requires 
%joe52
%using a large amount of states.} 
the use of many states.} 
as a consequence this approach does not seem to provide a satisfactory
explanation as to why people choose to cooperate.} 
%raf52: need more refs? maybe mention rubinstien, but he considers the IRPD

%Taking computation into account leads to a
%joe52
%In our framework, we can provide a 
{\rm By using our framework, we can provide a 
%taking computation into account leads to a
straightforward explanation.  
Consider the \emph{tit for tat} strategy, which proceeds as follows: a
player cooperates at the first round, and then at round $m+1$, does
whatever his opponent did at round $m$.  Thus, if the opponent cooperated
%joe55
%at the last step, then you reward him by continuing to cooperate; if he
%defected at the last step, you punish him by defecting.  If both players
at the previous round, then you reward him by continuing to cooperate; if he
defected at the previous round, you punish him by defecting.  If both players
play tit for tat, then they cooperate throughout the game.  Interestingly,
tit for tat does exceedingly well in FRPD tournaments, where computer
programs play each other \cite{Axelrod}.  }

%raf52:
%joe52
%{\rm Consider now a machine-game version of FRPD where at each round
{\rm Now consider a machine-game version of FRPD, where at each round
%raf107:
%the player receive as signal the move of the opponent in the previous
the players receive as a signal the move of the opponent in the previous
rounds before they choose their action. In such a game, tit for tat is a
simple program, which needs  
no
%very little 
%raf54:
%memory (i.e., the machine is stateless) and is deterministic.
%joe90*
%memory (i.e., the machine is stateless).
memory (i.e., the TM uses only one state).
%raf57: should we mention something about the fact that if a machine
%does not read its input it ``disappears'' 
%joe56: Sorry, Rafael, I didn't understand your point here.
%raf58*: we are assuming that memory is expensive, but what if i simply leave things on my input tape
%we are assuming that at each step the agent can only see the signal of that round. 
%joe55: removed paragraph break
%Suppose now that 
Suppose that 
we charge even a modest amount $\alpha$ for memory usage 
%joe54*: I thought we agreed that we didn't need to charge for
%randomness, right?
%raf54:indeed; but it seems to get a bit messy; in particular i am not
%sure it is true when u(D,D) = -1,-1. It seems that defecting with
%probability 1/3 gives you a decent chance at defecting in the last
%round without loosing enough  if you defect in earlier rounds. I
%changed u(D,D) to -3,-3 to prevent this. 
%joe55: I actually think this would have worked with -1, for N
%sufficiently large; see below.  But I certainly have no problem making
%the change.  I think that, in general, the result holds as long as the
%increase of (D,C) relative to (C,C) (in our case 2) is larger than 
%\delta(the decrease of (D,D) relative to (C,C)) (4 or 6, depending on
%whether you take the payoff to be -1 or -6. See below
%raf54:
%(i.e., stateful get a penalty of 
%joe90*: technically, all machines have at least one state in our framework
%(i.e., stateful machines get a penalty of 
(i.e., machines that use memory get a penalty of 
%at least $\alpha$, or more, whereas stateless machine get
%joe55: ``or more'' seems redundant
%at least $\alpha$, or more, whereas stateless machines get
%joe90
%at least $\alpha$, whereas stateless machines get
at least $\alpha$, whereas memoryless machines get
no penalty), 
%joe55
%and 
%joe54: put the 0.5 here
%consider the FRPD
%with a discount factor $\delta$, with $0 < \delta \le 1$, so that if the
%joe55
%that there is a 
that there is  
a discount factor $\delta$, with $0.5 < \delta < 1$, so that if the
player 
gets a reward of $r_m$ in round $m$, his total reward 
over the whole
$N$-round game 
%raf56:added
(excluding the complexity penalty)
is taken to be $\sum_{m=1}^N  \delta^m r_m$,
%joe55: added
that $\alpha \ge 2\delta^N$, and that $N > 2$.  
%joe54: removed paragraph break
%joe52: added \rm here (and a few other places
%{\rm 
In this
%raf52: discount > 0.5
%case, it is easy to see that for all $\delta < 1$, no matter what the
%joe54: now cut \delta
%case, it is easy to see that for all $0.5< \delta < 1$, no matter what the
%joe55: I guess it's not *so* easy
%raf55:unfortunately not..
%case, it is easy to see that, no matter what the
%cost of memory is, as long as it is positive, 
%for a sufficiently long game, 
case,
it will be a Nash equilibrium for both players to play
tit for tat.  
%joe55
%For the best response to tit-for-tat is to play
%raf56: the next sentence feels a bit out of place since after we
%consider the actual params above (but i kept it) 
Intuitively, no matter what the cost of memory is (as long as it is
positive), for a sufficiently long game, tit for tat is a Nash equilibrium.

%joe55: added paragraph break
%For the best response to tit-for-tat is to play
To see this, note that the best response to tit for tat is to play
%raf54: explained more as it is needed below
%tit-for-tat up to the last round, and then to defect 
tit for tat up to 
%joe55
but not including
the last round, and then to defect.
%joe55: cut this for now; I'll explain it below
%(note that any
%strategy that defects at a round $i<N$ does at least
%$6\delta^{i+1}-2\delta^{i}$ worse than tit-for-tat; when $\delta>0.5$
%this is positive). 
%(note that 
%joe52
%as $\delta > 0.5$, defecting at an earlier round has negative gain). 
%since $\delta > 0.5$, defecting at an earlier round leads to a worse
% outcome). 
But following
this strategy requires the player to keep track of the round number,
which requires the use of extra memory.  The extra gain of 2 achieved
%joe55
%by defecting at the last step, if 
%raf** remove if
by defecting at the last round
%raf56: remove if
%if
will not be worth the cost of keeping track of
%raf52: added
%the round number.
%raf54:
%the round number as long as $\alpha \geq 2\delta^N$; thus no randomized
the round number as long as $\alpha \geq 2\delta^N$; thus no 
%or 
%joe90
%stateful strategy can do better. 
strategy that use some memory can do better.
%raf54:
%Additionally, as long as $N>2$, no stateless deterministic machine do
%joe55
%We turn to consider (randomized) stateless strategies.
%As $N>2$ it follows that the probability that a stateless strategy
%joe90
%It remains to argue that no stateless strategy (even a randomized
%stateless strategy) can do better against tit for tat.  
It remains to argue that no memoryless strategy (even a randomized
memoryless strategy) can do better against tit for tat.  
%joe55: moved here, and expanded
Note that any strategy that defects for the first time at round $k<N$
does at least $6\delta^{k+1}-2\delta^{k}$ worse than tit for tat.  It
gains 2 at round 
$k$ (for a discounted utility of $2\delta^k$), but loses at least 6
%raf56
%relative to tit-for-tat in the next round, for a utility of
relative to tit for tat in the next round, for a discounted utility of
$6\delta^{k+1}$.
%raf56: added
% and from  
From that point on the best response is to either continue defecting (which at each round leads to a loss of 6), or cooperating until the last round and then defecting (which leads to an additional loss of 2 in round $k+1$, but a gain of 2 in round $N$). 
Thus, any strategy that defects at round $k<N$
does at least $6\delta^{k+1}-2\delta^{k}$ worse than tit for tat.

A strategy that defects at
the last round gains $2\delta^N$ relative to tit for tat.
%joe90
%Since $N>2$, the probability that a stateless strategy
Since $N>2$, the probability that a memoryless strategy 
%joe55
%defects before the last round is at least as high as the probability
%that it defects for the first time in the last round. It follows that
defects at round $N-1$ or earlier is at least as high as the probability
that it defects for the first time at round $N$. 
%raf56: added
(We require that $N>2$ to make sure that there exist some round $k<N$
where the strategy is run on input $C$.) 
It follows that
any stateless strategy that defects 
%joe55
%with non-zero probability $p$ (when playing against tit for tat) 
for the first time in the last round with probability $p$ 
in expectation gains at most $p
(2\delta^N-(6\delta^N - 2\delta^{N-1})) = p\delta^{N-1}(2 - 4\delta)$,
which is negative when $\delta > 0.5$. 
Thus, when $\alpha \geq 2\delta^N$, $N>2$, and $\delta>0.5$, tit for tat
is a Nash equilibrium in FRPD. 
%raf52: we can probably get around the cost of randomness by some extra
%calculations but this doesn't seem to give any further insights. 
%joe52: I agree
%raf52: at first I wanted to only let the cost always be $\alpha$, then
%we could probably provide a full characterization; but this doesn't
%seem like a reasonable model. 
%(On the other hand, if $\alpha < 2\delta^N$, or if $N \leq 2$, then
%always defecting will again be the only Nash equilibrium.) 
%joe52: the statement above is not so obvious to me.
%raf53: it requires the cost to always be \alpha
(However, also note that depending on the cost of memory,
%raf54
%or randomness, 
tit for tat may \emph{not} be a Nash equilibrium for sufficiently short
games.)} 
%raf54
%raf55: added para break

%joe37
%{\rm Note that the above argument extends to show that tit-for-tat is a
{\rm The argument above can be extended to show that tit for tat is a
Nash equilibrium even if there is also a charge for randomness or
computation, as long as there is no computational charge for machines as
``simple'' as tit for tat; this follows since adding such extra charges
can only make things worse for strategies other than tit for tat.} 
%raf55:added
%joe55*: Rafael, this is a little too mysterious.  In what way can you
%generalize the class of utility functions if we charge for randomness?
%raf56:removed
%{\rm (In fact, if we assume a small charge for the use of randomness,
%joe55
%the above argument can be simplified and continues to hold for a more
%the argument above continues to hold for a more
%general class of utility functions.)} 
%raf54:
%{\rm Note that even if only one player is computationally bounded and is
{\rm Also note that even if only one player is 
%raf54
%computationally bounded and is
%joe52
%raf54: back to old
charged for memory, and memory is free for the other player, then there
%charged for memory or randomization, and memory and randomization are
%free for the other player, then there 
is a Nash equilibrium where the bounded player plays tit for tat, while
the other player plays the best response of cooperating up to 
%joe55
but not including
the last
round of 
the game, and then defecting
%joe82
in the last round.
%joe52
%raf81*: I will add something about this being testable by showing them
%a clock. (and Rachel's comment about swimming)/ 
%joe82*: I think we should say that we can get the same result without
%using \delta (i.e., if \delta = 1), as long as the complexity grows
%unboundedly with N (the length of the game).  Added:
(A similar argument works without assuming a discount factor (or,
equivalently, taking $\delta=1$) if we assume that the cost of memory
increases unboundedly with $N$.)

%raf82: added, but you should probably rewrite it..
%joe74
%We mention that the assumption of a cost for remembering the round
%number could be testable by running experiments with a variant of the
%FRPD where the players have access to a counter annoucing how many
%rounds remain in the game. If players' behavoir changes with the
Note that the assumption of a cost for remembering the round
number can be tested by running experiments with a variant of the
FRPD where players have access to a counter showing how many
rounds remain in the game. If players' behavior changes in the
presence of such as counter, this could be interpreted as suggesting a
cognitive cost for remembering the round number. 
%joe74
%Indeed, as pointed out 
%to us by Rachel Cramption (who was previously a professional swimmer),
%in swimming contests, the swimmers have, and make use of, 
In fact, competitive swimmers have, and make use of, counters
telling them how many laps remain in the race.%
\footnote{We thank Rachel Kranton for this observation.}
%joe82: removed paragraph break
%
\qed}
\end{example}

%raf81: still needs fixing. 
%raf80:
%joe74: As I said, we may want to move this to a single-person decision
%theory paper eventually
%raf82: added example
%raf82: added
\begin{example}
[Biases in randomization]
\label{randombias:ex} 
{\em An interesting study by Walker and Wooders \citeyear{WW} shows that
%raf104:
%even in professional sports competition (when large sums of money are at
even in professional sports competitions (when large sums of money are at
play), players randomize badly. 
For instance, in tennis serving, we can view the server as choosing
%joe74
%between two actions: serving left or serving right. The receiver instead
%attempts to predict the action of the server. The (unique) minimax
between two actions: serving left or serving right, while the receiver 
attempts to predict the action of the server. 
%joe74*: I don't think it's true that they choose uniformly.  They
%choose so that they get the same expected utility from their choice,
%where the expected utility depends on the how good the server/receiver
%is at each serve and the prediction.  
%raf83: you are so right, i had only read the first two pages of the paper when i wrote this (and planned to read the rest later), so sorry about this!
%The (unique) minimax 
%equilibrium of this game requires both players to uniformly choose
%between their two actions.  
However, as observed by Walker and Wooders \citeyear{WW}, while players
%raf88:
%do indeed randomize, their choices are not made indepdently at each step.
do indeed randomize, their choices are not made independently at each step.
In particular, they tend to switch from one action to another too often.

%joe74
%If there is a cost for computing, these observations might not be
%surprising.   Assume that the server could actually randomize well if
%he wanted to, but that there is some, potentially small, cost for doing
%this; for instance, the server could use the nature around him
%to generate its randomness, but this might require a loss
%raf88:
%These observation can again be explained by assuming a cost for computation.
These observations can again be explained by assuming a cost for computation.
Suppose that the server could actually randomize well if he
wanted to, but that there is some (possibly small) cost for doing
%raf88:
%this.  For instance, the server make use of features of nature  
this.  For instance, the server may make use of features of nature  
%joe74
%its randomness, but this might require a loss
%in concentration. If the server indeed uses a weak method of
%randomization, the receiver might try to ``break'' it, but again this
%raf88:
to generate
%joe103
%its randomness, but this might entail a loss
her randomness, but this might entail a loss
in concentration.  Similarly, if the server indeed uses a weak method of
randomization, the receiver can try to ``break'' it, but this again
requires some computational effort (and again a loss of
concentration). 
%joe74
%So, the server might rationally assume that the receiver
%only uses a computationally-weak perdictor, and as such decides to only
%use a weak method of randomization that suffices to ``fool'' the weak
%class of predictors used by the receiver. 
So, under reasonable assumptions, there is an equilibrium where the
server uses only weak randomization, while the receiver uses only a
computationally-weak predictor (which can be fooled by the randomization
used by the server).

Indeed, the usage of such methods is prevalent in the computer
%joe74
%scientists litterature: the definition of \emph{pseudo-randomness}
%\cite{BM,Yao} recognizes that a sequence of numbers only needs to be
science literature: the definition of \emph{pseudo-randomness}
\cite{BM,Yao} recognizes that a sequence of numbers needs to be only
``random-looking'' with respect to a particular class of computationally
resource-bounded observers. It is also well-known that
computationally-weak classes of observers can be easily fooled. For
%joe74
%instance, by Hastad switching lemma \cite{Hastad}, it follows that a
%joe96: this is probably not the right level for economists
%instance, by Hastad's \citeyear{Hastad} switching lemma, it follows that a
%polynomial-size constant-depth circuit cannot distinguish a random
%sequence of bits whose parity is 1 from a random sequence with parity
%raf107
%example, a polynomial-size constant-depth circuit (a quite weak model
example, a polynomial-size constant-depth circuit (a weak model
of computation) cannot distinguish a random
sequence of bits whose parity is 1 from a random sequence with parity
0 \cite{Hastad}.

%raf102*:
Let us elaborate. We first recall the cryptographic
(complexity-theoretic) definition of a \emph{pseudorandom generator}
\cite{BM,Yao}. Intuitively, a pseudorandom generator (PRG) is a
deterministic function that takes a short random ``seed'', and expands
%joe96
%it to a longer string that ``looks'' random to a computationally
it to a longer string that ``looks random'' to a computationally
bounded observer. More precisely, call a TM $T$-bounded if 
%joe96
%its size and run-time is bounded by $T$; we now require that no
%joe103
%both its size and run-time are bounded by $T$; we now require that no
both its size and running time are bounded by $T$; we now require that no
$T$-bounded 
%joe96
%observer can ``guess the next bit'' of the sequence output by the PRG
observer can guess the next bit of the sequence output by the PRG
with probability significantly better than $1/2$, given any prefix of
%joe96
%it.\footnote{As shown by Yao \cite{Yao}, this condition is equivalent to
it.\footnote{As shown by Yao \citeyear{Yao}, this condition is equivalent to
saying that no computationally bounded ``distinguisher'' can tell apart
the output of the generator from a truly random string.}  

%joe96*: I wonder if this isn't too technical for economists
%joe96: it seems that you're using $A$ for a TM.  We've been using M up
%to now (and $A$ is the set of actions, so that may not be a good idea.
%I changed A to M, and made some other trivial changes
%Let $g: \bitset^{\ell_0} \rightarrow \bitset^{\ell}$, and let the
%boolean random variable ${\sf predict}^{g,i}_M(s,r) =1$ iff
%$A(g(s)_{0\rightarrow i}; r)=g(s)_{i+1}$, where $X_{0\rightarrow i}$
%denotes the first $i$ bits of $X$; that is ${\sf predict}^{g,i}_A(s,r) =
%1$ iff $A$ with randomness $r$ can guess the $i+1$'th bit of $g(s)$
%joe98:
%Let $g: \bitset^{\ell_0} \rightarrow \bitset^{\ell}$;
%intuitively, $g$ is a pseudorandom generator, and $\ell_0$ is much
More formally,  
%joe103
suppose that
$g: \bitset^{\ell_0} \rightarrow \bitset^{\ell}$.
(Intuitively, $g$ is a pseudorandom generator, and $\ell_0$ is much
smaller than $\ell$.)  Given a TM $M$, 
define the predicate ${\sf predict}^{g,i}_M: 
\bitset^{\ell_0} \times \bitset^{\infty} \rightarrow \{0,1\}$ by taking
%Boolean random variable 
${\sf predict}^{g,i}_M(s,r) =1$ iff
$M(g(s)_{0\rightarrow i}; r)=g(s)_{i+1}$, where $X_{0\rightarrow i}$
denotes the first $i$ bits of $X$; that is, ${\sf predict}^{g,i}_M(s,r) =
%joe96
%1$ iff $M$ with randomness $r$ can guess the ($i+1$)st bit of $g(s)$
%joe98: added commas
%1$ iff $M$ using the random string $r$ can guess the ($i+1$)st bit of $g(s)$
1$ iff $M$, using the random string, $r$ can guess the ($i+1$)st bit of $g(s)$
given only the first $i$ bits of $g(s)$. Let $\Pr_l$ be the uniform
distribution 
over $l$-bit strings.  

%joe96
%\BD We say that $g: \bitset^{\ell_0} \rightarrow \bitset^{\ell}$ is a
\BD The function $g: \bitset^{\ell_0} \rightarrow \bitset^{\ell}$ is a
$(T,\epsilon)$-{\em pseudorandom generator (PRG)} if $\ell >\ell_0$ and,
for all $T$-bounded TM machines $M$ and all $i \in
\{0,1,\ldots,\ell-1\}$,  
%joe96: you actually wrote $M$ two lines above, but A below
%$\Pr_{\ell_0}^+ [{\sf predict}^{g,i}_A = 1] \leq \frac{1}{2} + \epsilon$
$\Pr_{\ell_0}^+ [{\sf predict}^{g,i}_M = 1] \leq \frac{1}{2} + \epsilon$.
\ED
The existence of 
%joe97
efficient
pseudorandom generators has been established under a
%raf105: dof :-)
%variety of number-theoretic conjectures (e.g., the hardness dof factoring
variety of number-theoretic conjectures (e.g., the hardness of factoring
products of large primes, or the hardness of the discrete logarithm
%joe96: Rafael, can you send me the HILL reference?
%joe97*: Rafael, don't forget to send me the reference
%joe103*: Rafael, I still don't have the HILL reference
problem) \cite{BM,HILL}. 

Now consider the following simplified game representation of the
%joe96
%tennis serving scenario. The game consists of $\ell$ rounds where in
%round $i$, player 1 (the ``server'') picks a bit $x_i$ as its action,
tennis-serving scenario. The game consists of $\ell$ rounds, where in
round $i$, player 1 picks a bit $x_i$ as its action (intuitively, this
is meant to model the server choosing where to serve)
and player 2 may either pick a bit $g_{i}$ (``a guess for $x_i$'') or
output the empty string (for ``no guess''). The utilities are
specified as follows. For each round $i$ where player $2$ outputs a
%joe96
%correct guess (i.e., $g_i=x_i$), it gets a score of 1 whereas player 1
%gets a score of $-1$; when the guess is incorrect, player 1 instead
correct guess (i.e., $g_i=x_i$), player 2 gets a score of 1 and player 1
gets a score of $-1$; when the guess is incorrect, player 1 
gets a score of $1$ and player 2 a score of $-1$; if player 2 outputs
%joe96: I would think that player 1 should get more than 0 here,
%although I don't feel strongly about it
the empty string, both players get a score of 0. 
%joe98
(For simplicity, we are assuming that if player 2 does not try to guess
where player 1 will serve, then he will concentrate, and the players are
equally likely to do well.  In practice, the server has an advantage,
but this is irrelevant to the point we are trying to make.)
The final utility in
the game is the total score the players accumulate, with a penalty for
the computational resources used. For simplicity, player 1 receives a
``huge'' penalty $p>\ell$ if it uses more than $\ell_0$ random coins 
%joe98: slowed down
%and
%joe108
%(that is, if it looks at more than the first $\ell_0$ bits of the random
(i.e., if it looks at more than the first $\ell_0$ bits of the random
string $r$), and otherwise gets a penalty of 0;
player 2 receives 0 penalty if it ``does nothing'' (i.e., always
just outputs the empty string), a ``small'' penalty $p_0=\epsilon\ell+1$
if it uses a machine that does not always output the empty string, but
still is $T$-bounded, and a huge penalty $p$ if it uses a machine that
is not $T$-bounded. 
In such a scenario, it is easy to see that, assuming the existence of a
$(T',\epsilon)$-PRG $g:\bitset^{\ell_0}\rightarrow \bitset^{\ell}$,
where $T'$ is slightly larger than $T$ (the exact choice of $T$' is
explained below),
%$(T,\epsilon)$-PRG $g:\bitset^{\ell_0}\rightarrow \bitset^{\ell}$,
it is a NE for player $i$ to sample $\ell_0$ random bits $s$, compute
$x=g(s)$, and in round $i$ to output $x_i$, and for
%joe96
%player 2 to ``do nothing'':  By the security of the PRG $g$, we have that
player 2 to ``do nothing''.  By the security of the PRG $g$, we have that,
for every round $i$, player $2$ can guess $x_i$ with probability at most
$\frac{1}{2} + \epsilon$ if it uses a machine that is $T$-bounded, but
the potential $\epsilon$ advantage is eaten up by the ``small'' penalty
of $p_0$ for attempting a guess. 
%joe96*: I'm confused.  isn't player 2 already a TM violating the
%security of PRG.  Again, what's commented out below is my attempt to
%rewrite assuming we needed T'.
%joe97*: I started going back to my rewrite, but didn't finish, pending
%my questions above.
(The reason we require $g$ to be secure
for $T'>T$-bounded machine is a rather technical one:
there might be a small overhead in
turning a machine $M$ playing the role of player 2 in the game into a
machine $M'$ violating the security of the PRG.  
%joe98
Given an input $x$ of length $k$ and random string $r$, $M'$ simply
%joe103
%outputs what $M$ would output if $M$ had view $x$ (that if, in
%joe108
%outputs what $M$ would output if $M$ had view $x$ (that is, if, in
%consecutive rounds, $M$ saw $x_1, \ldots x_k$) and random string $r$.   
outputs what $M$ would output if $M$ had view $x$ (i.e., if, in
consecutive rounds, $M$ saw $x_1, \ldots, x_k$) and random string $r$.   
Although running $M$ takes at most $T$ steps, constructing the view may
take a few extra steps, so we need to have $T' > T$.)
%We can now explain the choice of $T'$.  For this to be a NE, it must be
%too computationally expensive for player 2 to predict effectively.
%Suppose that player 2 could predict effectively.  Then it must use a
%$T$-bounded TM $M$ to do so.  ...
%
%joe96*: I found this hard to parse.  What does it mean that player 1's
%strategy can be ``evaluated by T-bounded strategies?
%raf103*: recall that player 1 needs to run the PRG; the only thing i
%am saying here is that, if the PRG is efficiently computable (as they
%all are), then the above strategies are still a NE even if we add a
%huge penalty for using non-efficient strategies also to player 1.
%Note that the same strategies remain a NE if we add a ``huge'' penalty
%$p$ also to player 1 unless it uses a $T$-bounded strategy, 
%joe97*: So why do we even need to say ``as long as player 1's stratregy
%can be evaluated by a T-bounded strategy''?  Isn't that already
%implicit in the assumption that 1 is using a T-bounded strategy? 
%Rewrote in a way that captures what I think you are trying to say.  
%Note that the same strategies remain a NE if we  add a ``huge'' penalty
%$p$ to player 1 unless it uses a $T$-bounded strategy
%as long as
%player 1's strategy can be evaluated by $T$-bounded strategies.  Indeed,
%%joe96*: again, I don't understand this.  I doubt the economists will
%%either.  My sense is that we should just cut these last few sentences.
%%We're probably going into too much detail here.
%under standard cryptographic assumptions, we have PRGs that can be
%evaluated in time $\polylog T$.} 
As we mentioned, under standard cryptographic assumptions, there are
computationally efficient PRGs, so the same strategies 
remain a NE even if we  add a ``huge'' penalty
$p$ to player 1 unless it uses a $T$-bounded strategy.
}
\qed
\end{example}

%raf95: added this
%joe85: added label
%raf96: change positition of label
%\begin{example}\label{xam:biases}
\begin{example}
[Biases in decision theory]
\label{xam:biases}
{\em There are several examples of single-agent decision theory problem where
%joe90
%experimentally observed behavoir does not correspond to what is
experimentally observed behavior does not correspond to what is
predicted by standard models.  
For instance, psychologists have observed systematic biases in the way that
individuals update their beliefs as new information is received (see
\cite{mrabin98} for a survey). In particular, a
\emph{first-impressions-matter} bias has been observed: individuals put too
much weight on initial signals and less weight on later signals. As they
become more convinced that their beliefs are correct, many individuals
even seem to simply ignore all information once they reach a confidence
threshold. Other examples of such phenomena are \emph{belief polarization} 
%joe103
%\cite{LRL79}---that two people, hearing the same 
\cite{LRL79}---two people, hearing the same 
information (but with possibly different prior beliefs) can end up with
diametrically opposed conclusions, and the \emph{status quo} bias
\cite{SZ88}---people are much more likely to stick with what they
%joe85
%already have. In a companion paper \ref{HP10}, which applies our
%framework to decision theory, we show how these biases can be explain by
already have. In a companion paper \cite{HP10a}, where we apply our
framework to decision theory, we show how these biases can be explained by
considering computation as a costly resource.} 

{\em We here briefly show how the first-impressions-matter bias can be explained by considering a small cost for ``absorbing'' new information.
Consider the following simple game
%raf95
%raf96:added paren
%(which is very similar to the one studied in \cite{M98,W02}. 
%joe86: I don't think there's a unique game in either of these papers,
%so I removed ``the''
%(which is very similar to the one studied in \cite{M98,W02}). 
%joe104
%(which is very similar to one studied in \cite{M98,W02}). 
(which is very similar to one studied by Mullainathan \citeyear{M98} and
Wilson \citeyear{W02}). 
%joe85: I think you meant to comment out the next two lines
%raf96: i actually meant to comment out the line above and keep the two
%lines below, but not it is fine they way it is. 
%, which is very similar to the game studied in earlier work attempting
%to explain the first-impressions-matter bias (see e.g.,
%\cite{M98,W02}). 
%
The state of nature is a bit $b$ which is $1$
with probability $1/2$. An agent receives as his type a sequence of
independent samples $s_1, s_2,\ldots, s_n$ where $s_i = b$ with
probability $\rho > 1/2$. The samples corresponds to signals the agents
receive about $b$.  
An agent is supposed to output a guess $b'$ for the bit $b$. If the
guess is correct, he receives $1-mc$ as utility, and $-mc$ otherwise,
where $m$ is the number of bits of the type he read, and $c$ is the cost
of reading a single bit ($c$ should be thought of the cost of
absorbing/interpreting information). 
It seems
reasonable to assume that $c>0$; signals usually require some effort to
decode (such as reading a newspaper article, or attentively watching a
movie). 
If $c>0$, it easily follows by the Chernoff bound that after
reading a certain (fixed) number of signals $s_1, \ldots, s_i$, the
agents will have a sufficiently good estimate of $\rho$ that the
marginal cost of reading one extra signal $s_{i+1}$ is higher than the
expected gain of finding out the value of $s_{i+1}$. 
That is, after processing a certain number of signals, agents will
eventually  disregard all future signals and base their output 
guess only on the initial sequence.
%raf102*:
In fact, doing so stricly dominates reading more signals.
}
\qed
\end{example}

%raf56:
%joe56
%\subsubsection{Existence of Nash Equilibrum in General Classes of
%raf58: upgraded one level
%\subsubsection{A Sufficent Condition for the Existence of Nash
%raf60: conditions
%\subsection{A Sufficient Condition for the Existence of Nash
%raf90: upgraded
%\subsection{Sufficient Conditions for the Existence of Nash
\section{Sufficient Conditions for the Existence of Nash
Equilibrium 
%raf60: made shorter
%in Machine Games
}\label{sec:suff}
%joe37: slight rewriting
%We demonstrate the existence of Nash equilibria in some general classes
%of machine games. 
%We first focus on machine games where computation is free, i.e., the
%joe56: rewrote story
%Although, as Example~\ref{roshambo} shows, Nash equilibrium does not
%always exist in machine games, there are interesting classes of machine games
%where it is guaranteed to hold.  In this section, we characterize two 
%of them.  
%raf73:
As
Example~\ref{roshambo} shows, Nash equilibrium does not always exist in
machine games.  The complexity function in this example 
charged for randomization.  Our goal in this section is to show that
this is essentially the reason that Nash equilibrium did not exist; if
randomization were free (as it is, for example, in the model of
\cite{BKK07}), then Nash equilibrium would always exist.  

%joe56: more rewriting
%The first consists of  machine games where computation is free,
This result turns out to be surprisingly subtle.  To prove it, we first
consider machine games where \emph{all} computation is free,
that is, the utility of a player depends only on the type and action
profiles (and not the complexity profiles). 
%raf98*:
For ease of notation, we here restrict attention to Bayesian machine
games, but it can be easily seen that our results apply also to
extensive-form machine games. 
%joe56: cut all this
%but the players are still restricted to choosing randomized Turing
%machines.  
%%joe37
%%We next consider general machine 
%%games where computation is costly, but where pre-processing is free. 
%The second consists of general machine 
%games where computation is costly, but pre-processing is free. 
%
%\paragraph{Computationally cheap machine games}
%raf57: made uniform
%Let $\bar{\M}$ denote the set of randomized turing machine which (1)
%joe56: moved to where it's used
%Let $\M_B$ denote the set of randomized turing machine which (1)
%%joe37
%%have bounded output length (i.e., for each machine $M \in \bar{M}$ there
%%raf57:
%%have bounded output length (i.e., for each machine $M \in \bar{\M}$, there
%have $B$-bounded output length (i.e., for each machine $M \in \M_B$, there
%joe37*: T is overloaded (it's also the type space). More importantly, I
%think we want a uniform bound; see below
%exists some function $T(\cdot)$ such that for each $x\in\bitset^n$, the
%lenght of output of $M(x)$ is smaller than $T(n)$), and (2) terminate
%exists some function $B(\cdot)$ such that 
%for each $x\in\bitset^n$, the output of $M(x)$ has length 
%at most $B(n)$), and (2) terminate
%with probability 1. 
Formally, 
a machine game $G = ([m],\M,\Pr, T, \vec{\complex}, \vec{u})$ is 
%joe56
%said to be 
\emph{computationally cheap}
%joe103
%if $\vec{u}$ depends only on the type and action profiles, i.e., if 
if $\vec{u}$ depends only on the type and action profiles, that is, if 
there exists $\vec{u'}$ such that $\vec{u}(\vec{t},\vec{a},\vec{c}) =
\vec{u}'(\vec{t},\vec{a})$ for all $\vec{t},\vec{a},\vec{c}$.  

%joe37: slowed down a bit and added some intuition.  Next paragraph new.
%raf57: typo
%We would like to show that ever computationally cheap Bayesian
We would like to show that every computationally cheap Bayesian
machine game has a Nash equilibrium.
But this is too much to hope for.  The first problem is that the game
may have infinitely many possible actions, and may not be compact in any
reasonable topology.  This problem is easily solved; we will simply
require that the type space and the set of possible actions be finite.
%joe56: moved here and rewrote slightly
%joe86
%Given a \emph{bounding function} $B: \N \rightarrow \N$
%be a function,  a \emph{$B$-bounded Turing machine} $M$ is one 
Given a \emph{bounding function} $B: \N \rightarrow \N$,
a \emph{$B$-bounded Turing machine} $M$ is one 
%raf83: 
%that terminates with probability 1 on each input and, 
such that
for each $x\in\bitset^n$, the output of $M(x)$ has length 
%raf58:
%at most $B(n)$.  If we restrict to games with a finite type space where
at most $B(n)$.  If we restrict our attention to games with a finite type space where
only $B$-bounded machines can be used for some bounding function $B$,
%raf58:typo
%then we are guaranteed to only finitely many types and actions.
%joe86*: we do we have only finitely many actions?  It's true that for
%inputs of size n, for any fixed n, we have only finitely many actions.
%But the total space of actions is unbounded, isn't it?
%raf97*: hi joe, i am not sure to understand your comment. since the
%type space is finite,  
%there is a bound on $n$ and thus the actions space is finite too
then we are guaranteed to have only finitely many types and actions.
%joe87*: sorry; I'm just being slow
%that each pure strategy can involve only finitely many actions, but if
%we look over all strategies, there may be infinitely many actions, and
%a behavior strategy can randomize over infinitely many actions.  In any
%case, I don't see why a finite type space 

With this restriction, since we do not charge for computation in a
computationally cheap game, it may seem that this result should follow
trivially from the fact that every finite game has a Nash equilibrium.  
But this is false.  The problem is that the game itself might involve
%joe108
%non-computable features, so we cannot hope that that a Turing machine
non-computable features, so we cannot hope that a Turing machine
will be able to play a Nash equilibrium, even if it exists.

Recall that a real number $r$ is \emph{computable}
%raf56: ref to Minsky?
%joe37: actually, according to Wikipedia, computable numbers were
%introduced by Turing!
\cite{Turing37} if there exists a Turing machine that on input $n$
outputs a number $r'$ such that $|r-r'| < 2^{-n}$. 
%joe37
%We say that 
A game $G = ([m],\M,\Pr, T, \vec{\complexity}, \vec{u})$ is
\emph{computable} if (1) for every $\vec{t} \in T$, $\Pr[\vec{t}]$ is
computable, and (2) for every $\vec{t},\vec{a},\vec{c}$,
$u(\vec{t},\vec{a},\vec{c})$ is computable. 
%joe56: removed paragraph break
%joe37: added
%raf58: typo
%As we now, every computationally cheap \emph{computable} Bayesina
As we now show, every computationally cheap \emph{computable} Bayesian
machine game has a Nash equilibrium.  Even this result is not
immediate.  Although the game itself is computable, a priori, there may
not be a computable Nash equilibrium.  Moreover, even if there is, 
a Turing machine may not be able to simulate it.  Our proof deals with
both of these problems.

To deal with the first problem, we follow lines similar to those of
Lipton and Markakis \citeyear{LM04}, who used the Tarski-Seidenberg 
\emph{transfer principle} \cite{Tar1} to prove the existence
%raf58: changed slightly
%of \emph{algebraic} Nash equilibria in finite normal norm games. 
%joe2
%of \emph{algebraic} Nash equilibria in finite normal norm games with
of \emph{algebraic} Nash equilibria in finite normal-form games with
%joe92
%integer valued utilities.  
integer-valued utilities.  
%raf73:
(Similar techniques were also used by Papadimitriou and Roughgarden
%joe67
%\cite{PR08}.) 
%raf75:
%\citeyear{PR08}.) 
\citeyear{PR05}.
%joe105: How about putting Prasad here?  I feel like we should mention
%him someplace
Prasad \citeyear{Prasad09} proved that in finite normal-form games where
the utilities are computable, there exists a computable Nash equilibrium.)
We briefly review the relevant details here.

%joe37: moved this back, out of the proof
\BD
%joe37: slowed down a bit
An ordered field $R$ is a \emph{real closed field} if every positive
element $x$ is a square 
%joe37: added
(i.e., there exists a $y \in R$ such that $y^2 = x$), 
and every univariate polynomial of odd degree
%joe103
%with coefficients in $R$ has a root in $R$ 
with coefficients in $R$ has a root in $R$. 
\ED

%joe37: added
Of course, the real numbers are a real closed field.  It is not hard to
check that the computable numbers are a real closed field as well.

%joe37: added citation 
%joe85: to get numbering right
%\begin{theorem} [Tarski-Seidenberg \cite{Tar1}]
\begin{thm} [Tarski-Seidenberg \cite{Tar1}]
Let $R$ and $R'$ be real closed fields such that $R \subseteq R'$, and let
$\bar{P}$ be a finite set of (multivariate) polynomial inequalities with
coefficients in $R$. Then $\bar{P}$ has a solution in $R$ if and only if
it has a solution in $R'$. 
%joe85
%\end{theorem}
\end{thm}

%We say that a probability distribution $\Pr$ over $\bitset^n$ is
%\emph{samplable} if there exists a randomized Turing machine $M$ that
%samples the distribution $\Pr$ (i.e., for all $x \in \bitset^n$ it
%holds that the probability that $M$ outputs $x$ is $\Pr[x]$). 

With this background, we can state and prove the theorem.

\BT
\label{nashexist.the}
%joe37
%Let $G=([m], \bar{M},\Pr,\vec{\complex},\vec{u})$ be a computable
%bayesian machine game that is computationally cheap. Then there exits a
%Nash Equilibrium in $G$. 
%raf57:
%joe56*: We need to fix T and B up front
If $T$ is a finite type space, $B$ is a bounding function, $\M$ is a
%joe100
%set of $B$-bounded machines, then  
set of $B$-bounded machines, and
%$G=([m], \bar{M},\Pr,\vec{\complex},\vec{u})$ is a computable, 
%joe56
%$G=([m], \M_B,\Pr,\vec{\complex},\vec{u})$ is a computable, 
$G=([m], \M,\Pr,\vec{\complex},\vec{u})$ is a computable, 
%computationally cheap Bayesian machine game.
computationally cheap Bayesian machine game, 
%joe56
%with finite type space and $B(\cdot): \N \rightarrow \N$ be a function, 
then there exists a
Nash equilibrium in $G$.%
%joe104*
%raf108: i am not sure we should add this. actually prasad's result is
%very similar to lipton's result from 2004 that we reference; lipton
%shows that NE have algebraic probabilities if utilities are algebraic.
%above. our theorem apples to bayesian games (lipton/prasad only consider
%normal form games) and we here also consider costly computation.
%joe105: I feel like we should say something, in light of the blogs that
%you pointed out.  
%\footnote{Essentially the same result was obtained independently by
%Prasad \citeyear{Prasad09}.}
\ET

\begin{proof}
%joe37*: why is T finite?  Even if it is, it seems to me that there is
%raf58:by definition
%still a problem.  We know that each machine, is bounded, but the bounds
%may be different for each machine.  If that's the case, I see no reason
%why the set of actions (outputs) is finite.  I think we need a uniform
%bound.  
Note that since in $G$, (1) the type set is finite, (2) the machine set
contains only machines with bounded output length, and thus the action
set $A$ is finite, and (3) computation is free, 
%joe37
% $G$ can be viewed as a
%\emph{finite} Bayesian game $G'$ (with utility function $u'$).
%with the only restriction being that the strategies are required to be
%implementable by randomized Turing machines. 
there exists a finite (standard) Bayesian game $G'$ with the same type space,
action space, and utility functions as $G$.  
%joe37
%Recall that every finite Bayesian game has a Nash equilibrium \cite{}.
%raf56: is this also Nash's theorem or it came later?
%joe37: I'm not sure; let's punt on it.  This is such a well-known
%result that we can probably get away without a reference
Thus, $G'$ has a Nash equilibrium.

%joe37
%However, such a Nash equilibrium might not, a priori, be implementable
Although $G'$ has a Nash equilibrium, some equilibria of $G'$ might 
not be implementable
by a randomized Turing machine; indeed, Nash \citeyear{Nash51} showed
%joe37
%that there exist normal form games where all Nash equilibria are
that even if all utilities are rational, there exist normal-form games
where all Nash equilibria involve 
mixtures over actions with irrational probabilities. 
%raf56: was this in nash original paper?
%joe37: Lipton and Markakis reference a different Nash paper, so I did too
%As we show below, we can make use of the fact that the computable
%numbers form a a real closed field to deduce that there also exists a
%Nash equilibria which is computable by a randomized Turing
%machine.\footnote{Note that we do not prove that every Nash equilibrium
%can be computed by a randomized Turing machine, but merely that there
%exists one that is.}. Towards this goal, we rely on the transfer
%principle to ``transfer'' a real-valued equilibrium into a computable
%one. 
To deal with this problem we use the transfer principle.

%joe37
%In our context, we let $R'$ denote the real numbers and $R$ the
Let $R'$ be the real numbers and $R$ be the
computable numbers.  Clearly $R \subset R'$.
%joe37: let's credit Lipton-Markakis for this idea
We use the approach of Lipton and Markakis to show that a Nash
equilibrium in $G'$ must be the
%The idea will be to show that Nash equilibria in $G'$ can be seen as a 
solution to a set of polynomial inequalities with
coefficients in $R$ (i.e., with computable coefficients). 
%Then, by combining the Theorem by Tarski-Siedenberg with the fact that
%all finite Bayesian games have a Nash equilibrium, we conclude that
%there is also a 
%joe57
%Then, by combining the Tarski-Siedenberg transfer principle with the
Then, by combining the Tarski-Seidenberg transfer principle with the
fact that $G'$ has a Nash equilibrium,
%Nash equilibrium where the probability of each action, given a type, is
%computatble. We next show that if each such probability is computable,
%there also exists a Turing machine which samples according to that
%distribution, while terminating with probability 1. 
it follows that there is a computable Nash equilibrium.

%joe37
%To see that a Nash equilibria in Bayesian games can be viewed as a
%solution to a set of polynomial inequalities, let $\sigma_i(t_i,a_i)$
%denote the probability with which player $i$ chooses action $a_i$ given
%the type $t_i$, in a presumed equilibrium. 
The polynomial equations characterizing the Nash equilibria of $G'$ are
easy to characterize.
%joe37: I don't think we need this; we just use the definition of Nash
%equilibrium.  
%Recall that in a Bayesian game, for each type 
%the Nash equilibrium strategy must be a
%best response for each player $i$ also conditioned on the type $t_i$ for
%each $t_i$. Also, recall that without loss of generality it is
%sufficient to consider ``deviations'' only with respect to pure
%strategies. 
%It follows that $\vec{\sigma}$ is a Nash equilbrium if and only if
%(1) for each $i, t_i, a_i$, $\sigma(t_i,a_i) \geq 0$, (2) for each $i,
%t_i$, $\sum_{a_i \in A_i} \sigma(t_i,a_i) = 1$, and (3) for each
%$i,t_i, a'_i$, 
By definition, $\vec{\sigma}$ is a Nash equilibrium of $G'$ if and only if
(1) for each player $i$, each type $t_i \in T_i$, and  $a_i \in A_i$,
$\sigma(t_i,a_i) \geq 
0$, (2) for each player $i$ and 
$t_i$, $\sum_{a_i \in A_i} \sigma(t_i,a_i) = 1$, and (3) for each player
$i$ $t_i \in T$, and action $a'_i \in A$, 
%joe37
%$$\sum_{\vec{t}_{-i} \in T_{-i}} \sum_{\vec{a} \in A} \Pr [t]
$$\sum_{\vec{t}_{-i} \in T_{-i}} \sum_{\vec{a} \in A} \Pr (\vec{t})
u'_i(\vec{t},\vec{a}) \prod_{j\in [m]} \sigma_j(t_j,a_j) \geq  
%joe37
%\sum_{\vec{t}_{-i} \in T_{-i}} \sum_{\vec{a}_{-i} \in A_{-i}} \Pr [t]
\sum_{\vec{t}_{-i} \in T_{-i}} \sum_{\vec{a}_{-i} \in A_{-i}} \Pr (\vec{t})
u'_i(\vec{t},(a'_i, \vec{a}_{-i})) \prod_{j\in [m]\backslash i}
\sigma_j(t_j,a_j). 
$$
%joe37
Here we are using the fact that a Nash equilibrium must continue to be a
Nash equilibrium conditional on each type.

%joe37
Let $P$ be the set of polynomial equations that result by replacing 
$\sigma_j(t_j,a_j)$ by the variable  $x_{j,t_j,a_j}$.  
Since both the type set and action set is finite, and since both the
type distribution and utilities are computable, this is a finite set of
polynomial inequalities with computable coefficients,
%joe37
whose solutions are the Nash equilibria of $G'$.
%joe37
%It follows that
It now follows from the transfer theorem that
$G'$ has a Nash equilibrium where all the probabilities
$\sigma_i(t_i,a_i)$ are computable. 

%joe37
%It only remains to show that this equilibrium can be implemented by a
It remains only to show that this equilibrium can be implemented by a
randomized Turing machine. We show that, for each player $i$, and each
type $t_i$,  
there exists a randomized machine that samples according to the
distribution $\sigma_i(t_i)$; since the type set is finite, this
%joe37
%concludes that there exists a machine that implements the strategy
implies that there exists a machine that implements the strategy
$\sigma_i$. 

%joe37: changed superscripts to subscripts.
%Let $a^1, \ldots, a^N$ denote the actions for player $i$, and let
Let $a_1, \ldots, a_N$ denote the actions for player $i$, and let
$0=s_0\leq s_1 \leq \ldots \leq s_N=1$ be a sequence of numbers such
that $\sigma_i(t_i,a_j) = s_j - s_{j-1}$. Note that since $\sigma$ is
%joe37
%computable, the sequence $s_0, \ldots, s_N$ is computable too.  
computable, each number $s_j$ is computable too.  
%joe37
%That means that there exists a machine which on input $n$ computes an
That means that there exists a machine that, on input $n$, computes an
approximation $\tilde{s}_{n_j}$ to $s_j$, such that $\tilde{s}_{n_j} -
%raf58*: is the 2_{-n} notation standard?
%joe67: I think it's a typo, and was meant to be 2^{-n}$; I changed it
%everywhere.  I'm not sure how it happened!
%raf73*: i am still confused by the notation 2_-n (shouldn't we have 2^-n)?
2^{-n} \leq s_j \leq 
\tilde{s}_{n_j} + 2^{-n}$. 
%joe103
%Consider now the machine $M_i^{t_i}$ that proceeds 
Now consider the machine $M_i^{t_i}$ that proceeds 
%joe37
%in iterations 
as follows.  
%joe37: rewrote
The machine constructs a binary decimal $.r_1r_2r_3\ldots$ 
bit by bit.  After the $n$th step of the construction, the machine
%joe103*: Rafael, I rewrote this again.  It seems to me that every
%finite prefix is in a unique interval, and that's not what we meant.
%checks if the decimal constructed thus far ($.r_1\ldots r_n$) is
%guaranteed to be a unique interval $(s_k,s_{k+1}]$.  (Since $s_0,
checks if the prefix of the decimal constructed thus far ($.r_1\ldots
r_n$) already determines in which interval  $(s_k,s_{k+1}]$ the full
decimal is in.  (Since $s_0, 
\ldots, s_N$
are computable, it can do this by approximating each one to within
$2^{-n}$.)  With probability 1, after a finite number of steps, the
decimal expansion will be known to lie in a unique interval
$(s_k,s_{k+1}]$.  When this happens, action $a_k$ is performed.
%$r_1$ at random.  If there is a unique action $a_k$ whose 
%In iteration $n$, it samples a random bit $r_n$---the $n$'th
%digit in the binary expansion of a number $r_n$ between 0 and 1---and
%finds the set of $k$'s such that $[r_n, r_n+2^{-n}]$ and
%$[\tilde{s}_{k-1}_n - 2^{-n}, \tilde{s}_k_n + 2^{-n}]$ intersect. If the
%set contains more than one element it continues sampling another bit
%(increasing $n$); otherwise, if there is only one remaining $k$, it
%outputs $a_k$. 
%
%First note that by construction it follows that there exists at least
%one $k$ at each iteratation $n$ (since $s_0 = 0$, $s_N=1$ and
%$\tilde{s}_j_n$ is a $2^{-n}$-approximation of $s_j$). Furthermore,
%assuming that $M_i^{t_i}$ always terminates, it immediately follows
%that it samples according to $\sigma_i(t_i)$. It only remains to show
%that $M_i^{t_i}$ terminates with probability 1. 
%Note that $M_i^{t_i}$ continues in iteration $n$ 
%if there exists some $k$ such that both $k$ and $k+1$ survive in
%interation $n$, i.e., $[r_n, r_n+2^{-n}]$ and $[\tilde{s}^{k}_n -
%2^{-n}, \tilde{s}^k_n + 2^{-n}]$ intersect. For any fix $k$, it follows
%that the probability of both $k$ and $k+1$ surviving after $n$
%iterations is $O(2^{-n})$; we conclude (by the union bound) that the
%probability of (at least) two $k$'s surviving after $n$ iterations is
%smaller than $O(|A_i|2^{-n})$. 
\end{proof}

%joe56*: rewrote completely.  Let me know what you think
\commentout{
\paragraph{Games with cheap pre-processing}
We next consider games where computation is costly but pre-processing is
free---namely, the players are allowed to perform any arbitrary
computation before looking at their type; thereafter computation becomes
costly. 
%joe37*: Rafael, this needs more discussion.
%raf57: left as is now, we can go back to this after your general
%discussion about what NE in machine games means.  
Cheap pre-processing can be viewed as way to model the ``populational"
aspect of a strategy: a player's ``cultural'' heritage is cheap, but the
actual computation it has to perform costly. 

%joe37*: I had a lot of trouble with this description and the proof.
%Let's talk about it.
%raf57:
Let $\M'$ be a finite set of machines with finite running-time.
In a \emph{game with cheap pre-processing}, we consider a set of
%joe37
%machines $M$ (which terminate with probability 1) and that on input the
machines $M$ that terminate with probability 1 and that, on input the 
%raf57:
random tape $r$ and  
type $t$, proceed in two stages: 
%raf57:
%in the first stage $M$ performs some
in the first stage $M(r,t)$ performs some
computation---without reading the type $t$---and outputs $M' = M(r_1)$
where $r_1$ is a prefix of $r$ and  
%raf57:
%a finite machine $M'$ with finite running-time; in the second stage $M'$
$M'\in \M'$; in the second stage $M'$
is executed on input the 
the remaining part $r_2$ of the random tape $r$ and the
type $t$. The complexity of such a machine $M$
%raf57
in a view $(r_1,r_2);t$ 
%is only a function of $M'$, $t$ and the random tapes used by $M'$. 
is only a function of $M'=M(r_1)$, $t$ and the random tape $r_2$ used by
$M'$. 
}
%joe56: \end{commentout}

%joe56*: new material
We now want to prove that a Nash equilibrium is guaranteed to exist
provided that randomization is free.  Thus, we assume that we start with
a 
%raf58: added
finite
set $\M_0$ of \emph{deterministic} Turing machines and a finite set
%joe82: missing period
$T$ of types. 
%raf83:
%(and continue to assume that all the machines in $\M_0$ terminate).  
$\M$ is the
\emph{computable convex closure of $\M_0$} if $\M$ consists of machines
$M$ that, on input $(t,r)$, first perform some computation that depends
only on the random string $r$ and not on the type $t$, 
%raf58:
that with probability 1,
after some finite
time and after reading a finite prefix $r_1$ of $r$, choose a machine $M' \in
\M_0$, then run $M'$ on input $(t,r_2)$, where $r_2$ is the remaining
part of the random tape $r$. 
%raf105: added
The only output of $M$ is the final output of $M'$.
%joe82*: we should say that this means that the complexity of a convex
%combination of machines is the convex combination of their complexities
%raf58: typo
%type $t$.  
Intuitively, $M$ is randomizing
over the machines in $\M_0$.  
%raf58: we do need it
%(Although we do not need this
%fact, it 
%raf85: should we remove this now?
%joe76: It seems that we need this to appeal to the previous result,
%don't we?
%raf86*: (i think) the only reason for B-boundedness was to ensure that
%the action space is finite.  
%It
%joe103: typo, I presume
%Since machines in $\M$ are deterministic, it
Since machines in $\M_0$ are deterministic
%raf105: added to clarify
(and the set $\M_0$ is finite)%
, it
is easy to see that there must be some $B$ such that all the
machines in $\M$ are $B$-bounded.
%joe103*: Rafael, I'm not convinced.  Suppose that M writes a 1 at each
%step until it reads 3 consecutive 1s on the random tape, and then stops
%(or runs some arbitrary machine M' in \M_0; it doesn't matter.  So,
%with probability 1, M does choose a machine in finite time, but there
%is no bound B(n) on the length of the output of M as a function of the
%length of its input.  Indeed, the output is independent of the input.
%Am I misisng something here?
%raf105*: recall that M's only outut is the output of the machine
%M'\in M_0 selected 
%)  
\emph{Randomization is free} in a machine game $G = ([m],\M,\Pr, T,
\vec{\complexity}, \vec{u})$ where $\M$ is the computable convex closure
of $\M_0$ if $\complexity_i(M,t,r)$ is $\complexity_i(M',t,r_2)$ (using
the notation from above).  
%raf58*:i don't understand the statement below (removed it for now)
%That is, the complexity of $M$ is the obvious
%convex combination of the complexities of the machines in $\M_0$ over
%which it is randomizing.

\BT
\label{nashpreproc.the}
%joe37
%Let $G=([m], \M,\Pr,\vec{\complex},\vec{u})$ be a computable Bayesian
%machine game with cheap pre-processing. Then there exits a Nash
%joe56: again, we need the types
%If $G=([m], \M,\Pr,\vec{\complex},\vec{u})$ if a computable Bayesian
%machine game with cheap pre-processing and finite type space, 
If $\M$ is the computable convex closure of some finite set $\M_0$ of
deterministic Turing machines, $T$ is a finite type space, and 
$G=([m], \M,\T, Pr,\vec{\complex},\vec{u})$ is a game where
randomization is free,
then there exists a Nash
equilibrium in $G$. 
\ET
\begin{sketch}
%joe56: rewrote
%First note that every machine game with a finite machine set, where each
%machine terminates after a finite number of steps, can be viewed as a
%%joe56
%%finite normal form game---i.e., not even a Bayesian game---where the
%finite normal-form game (i.e., a Bayesian game where the type space is a
%singleton), where the
%action set is the set of machines. 
First consider the normal-form game where the agents choose a machine in
$\M_0$, and the payoff of player $i$ if $\vec{M}$ is chosen is the
expected payoff in $G$ (where the expectation is taken with respect to
the probability $\Pr$ on $T$).  
By Nash's theorem, it follows that
there exists a mixed strategy Nash equilibrium in this game. Using the
%joe56
%same argument as in the proof of theorem \ref{nashexist.the}, it follows
same argument as in the proof of Theorem \ref{nashexist.the}, it follows
%joe56
%that there exists a randomized Turing machine that samples according to
that there exists a machine in $\M$ that samples according to
the mixed distribution over machines, as long as the game is computable
(i.e., the type distribution and utilities are computable)
%raf57:
and the type and action spaces finite. 
%raf58:added
%raf83: i am happy to leave the B-bounded thing here, as we mentioned
%above that the machines trivially are B-bounded, and we used B-bounded
%in the earlier proof that we refer to. but i am of also very happy to
%change the reason to determinism if you prefer it. 
(The fact that the action space is finite again follows from the fact that type
%joe77
%space is finite and that 
space and $\M_0$ are finite and that 
%joe75: I prefer deterministic (it's also shorter) 
%raf86*: we need both deterministic and that $\M_0$ is finite.
%joe77: I think the change I made above should take care of this
%raf85: should we remove the B-bounded sentence above?
%there exist some $B$ such that all machine in $\M$ are $B$-bounded.) 
%joe77
%all machines in $\M$ are deterministic.)
all machines in $\M_0$ are deterministic.)
%joe56
%This
%concludes that there exists a Nash equilibrium in every computable
%Bayesian machine game with cheap pre-processing
%%raf57:
%and finite type space.
%raf58:typo
%The desired rseult follows.
The desired result follows.
\end{sketch}

%joe56: the next paragraph should be polished a bit
%raf58: i actually like it a lot.
We remark that if we take Aumann's \citeyear{Aumann87} view of a
mixed-strategy equilibrium as representing an equilibrium in players'
beliefs---that is, each player is actually 
using a deterministic strategy in $\M_0$,  and the probability that
player $i$ plays a strategy $M' \in \M_0$ in equilibrium represents 
%joe103
%all
the other players' beliefs about the probability that $M'$ will be
played---then we can justify randomization being free, since players are
not actually randomizing.
%raf58:added
%joe57
%The fact that the randomization is computable here represent that 
%players' beliefs are computable (such a requirement was advocated by
%Megiddo \cite{Med2}) 
%and that the population players are chosen from, can be sampled by a
The fact that the randomization is computable here amounts to the
assumption that players' beliefs are computable (a requirement advocated by
Megiddo \citeyear{Meg2}) 
and that the population players are chosen from can be sampled by a
Turing machine. 
More generally, there may be settings where
randomization devices are essentially freely available (although, even
then, it may not be so easy to create an arbitrary computable
distribution).  

%raf57*: need to check the details below, but approx this should hold.
%joe56
%The above result shows the existence of Nash equilibria when
%preprocessing is free. As we argue below, if we settle for an
%$\epsilon$-Nash equilibrium it is enough to assume that
Theorem~\ref{nashpreproc.the} shows that if randomization is free, a
Nash equilibrium in machine games is guaranteed to exist.  We can
generalize this argument to show that, to guarantee the existence of 
%joe103
%$\epsilon$-Nash equilibrium (for arbitrarily small $\epsilon$) it is
$\epsilon$-Nash equilibrium (for arbitrarily small $\epsilon$), it is
enough to assume that 
%raf58: preproc->rand
%``polynomial-time'' preprocessing is free.  
``polynomial-time'' randomization is free.  
%raf58:
%Lipton and Markakis \citeyear{LM04} show
Lipton, Markakis and Mehta \citeyear{LMM03} show
%joe56: we should either cut this or expand, perhaps with a reference to
%Alon and Spencer
%using probabilistic method, 
that 
every finite game with action space $A$ has an $\epsilon$-Nash equilibrium
%joe103
%with support on only $poly(\log |A| + 1/\epsilon)$ actions; furthermore
with support on only $poly(\log |A| + 1/\epsilon)$ actions; furthermore,
the probability of each action is a rational number of length $poly(\log
|A| + 1/\epsilon)$. In our setting, it follows that there exists an
%joe56*: Do we need circuits rather than Turing machines?  If so, doesn't
%that cause a problem?
%raf58: no, changing to TM
%$\epsilon$-NE where the preprocessing can be computed by a circuit of
%raf58: preproc->rand
%$\epsilon$-Nash equilibrium where the preprocessing can be computed by a
$\epsilon$-Nash equilibrium where the randomization can be computed by a
%raf58:
%circuit of 
Turing machine with size and running-time bounded by
size $O(\log |\M'| + 1/\epsilon)$. 
%raf56**: By the probabilistic method, we can also show that there
%exists a epsilon-Nash in games where O(log |M'| + 1/\eps) computation
%is free. (I think something similar was used by Lipton to show that
%eps-Nash can be found in Qpoly time). If assuming that M' only contains
%polynomial-sized Turing machines with polynomial running-time, then we
%get that there always exists a 1/poly-NE if assuming that poly
%pre-processing is free. Is this worth adding?  
%joe37*: we should think about our intended audience here.  I think it
%depends in part on how complicated the proof is.  Let's talk about this too.
%joe56
We omit the details here.

\section{Conclusion}\label{sec:conclusion}
%raf95*: comment below is still valid.
%raf90*: this section is piddly now; let's discuss it.
%joe85*: I agree!
We have defined a general approach to taking computation into account in
%raf107: softened the claim
%game theory that subsumes previous approaches.
game theory that generalizes previous approaches.
%raf107:
This opens the door to a number of exciting
research directions. We
briefly describe a few here:
\begin{itemize}
%joe89*: switched the order of items
\item 
In our framework we are assuming that the agents understand the costs
associated with each Turing machine.  That is, they do not have to do
any ``exploration'' to compute 
the costs.  As mentioned, we can model uncertainty regarding complexity
%joe103
%by letting the complexity function take also the state of nature as
by letting the complexity function also take the state of nature as
input. 
However, even if we do this, we are assuming that agents can
compute the probability of (or, at least, are willing to assign a
probability to) events like ``TM $M$ will halt in 10,000 steps'' or
``the output of TM $M$ solves the problem I am interested in on this
%joe89
%input''.  But, calculating such probabilities itself involves
input''.  But calculating such probabilities itself involves
computation, which might be costly.
%raf100*: I beleive the above can be solved by considering kripke
%structures where we charge for beliefs and perhaps also the
%representation of the structure. 
%joe89*: This might be an interesting direction to go, one related to
%your notion of ``comptuational knowledge''.  I would be *very*
%interested in discussing this further.
%joe89: removed paragraph break, and rewrote
%
%In addition, we do not charge the players for
Similarly, we do not charge the players for
computing which machine is the best one to use in a given setting, 
%joe89
%or even for computing the utilities of certain machine.
or for computing the utility of a given machine profile.
%raf100* below is from DT paper.
%joe89: cut this sentence.  I think it's too much
%Of course, calculating utilities may involve computation;
%although the utility was easy to compute in the simple examples given in
%this paper,  this is certainly not the case in general.  
It would be  
relatively straightforward to extend our framework so that the TMs
computed probabilities and utilities, as well as actions;
%joe89
this would allow us to ``charge'' for the cost of computing
probabilities and utilities.
However, once we do this, we need to think about what counts as an
``optimal'' decision if the 
%raf107:
%DM 
player
does not have a probability and utility,
or has a probability only on a coarse space.

%joe89*
%An alternative approach might be to allow the set of TMs that the DM
%considers possible to increase (at some computational cost), but assume
%that DM has all the relevant probabilistic information about the TMs
%that it can choose among.  
A related problem is that we have assumed that the players have all options
available ``for free''.  That is, the players choose among a set of TMs
which is given \emph{ex ante}, and does not change.  But, in practice,
it may be difficult to generate alternatives.  (Think of a chess player
trying to generate interesting lines of play, for example.)  Again, we
can imagine charging for generating such alternatives.  One approach we
are currently exploring for 
dealing with this 
%raf101: 
%joe90: I actually meant ``complex'' here, but cluster is perhaps better 
%complex 
%type
%of problems is to assume that the DM starts
cluster
of problems is to assume that 
%raf107
each player
%the  DM 
starts
with a small 
set of TMs that he understands (by which we mean that he understands the
behavior of the TM, and has a good estimate of its running time).  One of
the TMs he can choose involves ``thinking'';  the result of such
thinking is to perhaps produce further alternatives (i.e., more TMs that
he can choose among), or to give a more refined or accurate estimate of
the probabilities, utilities, and complexities for the TMs that he is
currently aware of.  If the player believes that he has already
generated enough good alternatives, he may decide to stop thinking, and
act.  But, in general, he can go back to thinking at any time.  In any
case, in such a framework, it makes sense to talk about a player making
a best response at all times, even as he is generating for alternatives.
These ideas are clearly related to work on awareness in games (see, e.g.,
\cite{HMS07,HR06}); we are planning to explore the connection.

%raf100*: 
%joe85*: we now say something in HP10 too!  We should figure out what to
%say here about that. 
%raf100*: took from DT paper (it is not an open question but is it
%certainly a research direction opened up by this paper) did i copy too
%much? 

\item In a companion paper \cite{HP10a}, we apply our framework to
decision theory. 
%joe89: added next line, shrunk the rest
%As an application, we use the framework  to define a
We show that the approach can be used to explain the status quo bias
\cite{SZ88}, belief polarization \cite{LRL79}, and other biases in
information processing.  Moreover, 
we use the framework  to define  two extensions of the standard notion
of \emph{value of information}: \emph{value of computational
information} and \emph{value of conversation}.   Recall   
%joe89*: shortened significantly
%that the \emph{value of information}, a standard notion in decision
%analysis, is meant to be a measure of how much a decision maker (DM)
%should be  willing to pay to receive new information.   
%The idea is that, before receiving the
%information, the DM has a probability on a set of relevant events and
%chooses the action that maximizes his expected utility, given that
%probability.  If he receives new information, he can update his
%probabilities (by conditioning on the information) and again choose the
%action that maximizes expected utility.  The difference between the
%expected utility before and after receiving the information is the value
%of the information.
that value of information
is meant to be a measure of how much a decision maker (DM)
should be  willing to pay to receive new information.   
In many cases, a DM seems to be receiving valuable information that
is not about what seem to be the relevant events.  This means that we
cannot do a value of information calculation, at least not in the
obvious way.  For example, suppose that the DM is interested in
learning a secret, which we assume for simplicity is a number between 1
and 1000.  A priori, suppose that the DM takes each number to be
equally likely, and so has probability $.001$.   
Learning the secret has utility, say, \$1,000,000; not learning it has
utility 0.  The number is locked in
a safe, whose combination is a 40-digit binary numbers.  
What is the value to the DM of learning the first 20 digits of the
combination?  As far as value of information goes, it seems that the
value is 0.  The events relevant to the expected utility are the possible
values of the secret;
learning the combination does not change the probabilities of
the numbers at all.  This is true even if we put the possible
combinations of the lock into the sample space.  
On the other hand, it is clear that people may well be willing to pay
for learning the first 20 digits.  It converts an infeasible
problem (trying $2^{40}$ combinations by brute force) to a feasible
problem (trying $2^{20}$ combinations).  
%joe89*
Because we charge for computation, it becomes straightforward to define
a notion of value of computational information that captures the value
of learning the first 20 digits.%
\footnote{Our notion of value of computational information is related
to, but not quite the same as, the notion of \emph{value of computation}
%joe89
%introduced by Horvitz \citeyear{Hor87,Hor01}.}
introduced by Horvitz \citeyear{Hor87,Hor01}; see \cite{HP10a} for
further discussion.}
%joe89: removed paragraph break
%
%joe89*: shortened
%raf100: removed
%Although this example is clearly contrived, there are many far more
%realistic situations where people are clearly 
%willing to pay for information to improve computation.  For example,
%companies pay to learn about a manufacturing process that will speed up
%production; people buy books on speedreading; and faster algorithms for
%search clearly are considered valuable.
%joe89*: 
%We show that we can use our computational framework to make the notion
%%raf100:
%%of \emph{value of computational information} precise, in a way that
%of \emph{value of computational information} precise.%
%, in a way that
%makes it a special case of value of information.%
%joe89*: The more I look at it, I wonder if we shouldn't talk about
%''(computational) value of conversation'', rather than ``value of
%(computational) conversation''.  The conversation has nothing to do
%with computation, after all.  ``Value of (computational) information''
%is not so bad, because the DM is getting information that helps in
%computation.  What do you think, Rafael?
%We further extend these notions to consider a notion of \emph{value of
%(computational) conversation, where the DM can communicate
%interactively 
The notion of value of
(computational) conversation further extends these ideas by allowing a
DM to interact
with an informed observer before making a decision (as
opposed to just getting some information).
%raf100:
%joe89*: I think this is too much detail for this paper
%Interestingly, whereas when computation is free, the value of conversion
%can be viewed as a special case of value of information, this is no
%longer the case when computation is costly. 
%joe89*: added, to make it seem like there is new stuff to be done
In making these definitions, we assumed, as we did in this paper, that
all the relevant probabilities and utilities were given.  It seems
interesting to explore how things change in this setting if we charge
for computing the probabilities and utilities.

\item In our framework, the complexity profile depends only on the
strategy profile of the players and the inputs to the machines.  
%raf95: the above sentence it not entirely accurate since in a nash we
%could charge for our beliefs in a specific way as player $i$ utility
%depends on the complexity of $-i$, and the strategies of $-i$ is the
%belief of $i$. 
%joe103
%We could extend the framework to also require that players' beliefs are
We could extend the framework to also require that players' beliefs be
computable by Turing machines, and additionally charge for the
complexity of those beliefs. For instance, a player might wish to play a
strategy that can be justified by a ``simple'' belief: Consider a
%joe90
%politician that needs to have a succint explanation for his actions. 
politician that needs to have a succinct explanation for his actions. 
%joe85*: I think the really interesting question is extending
%computation costs to computing the probability, utility, and what
%strategies to use.  We definitely need to say something more about
%this, especially in the context of extensive-form games (where more
%computation might make more TMs available).
%raf95: added
%joe103*: moved this back, and folded it into the rationalizability bullet
%%raf100: reordered
%\item As we have seen, Nash equilibria do not always exist
%in games with computation.  
%This leaves
%open the question of what the appropriate solution 
%concept s.
%%raf90*: needs more here, i would like to argue that it might not be a
%%bad thing that they don't exist, for instance in roshambo we might
%%want ``chaos'' to be the right answer? 
%joe103: rewrote slightly
%\item A natural next step would be to provide a computational
%definition of rationalizability.
\item We have focused here on Nash equilibrium.  But, as we have seen,
Nash equilibria do not always exist.  This makes the question of what is
%raf107
%the ``right'' solution concept for comutational games of particlar
%joe108
% the ``right'' solution concept for computational games of particlar
 the ``right'' solution concept for computational games of particular
%joe104
%intereset.   We could certainly consider
interest.   We could certainly consider
other solution concepts, such as rationalizability, in the computational
context.  
%joe85*
%joe103
%Here we might want to take seriously the cost of computing the
With rationalizability,  we might want to take seriously the cost of
computing the 
beliefs that rationalize the choice of a strategy.

%raf100:
\item It would be interesting to provide epistemic characterization of  
various solution concepts for machine games. It seems relatively
straightforward to define Kripke structures for machine games that model
agents' beliefs about their own and others complexities and utilities,
and their belief about other agents' beliefs about complexity and
%joe89
%utilitity (as well as their beliefs about beliefs etc.)  
utility (as well as their beliefs about beliefs etc.).
%joe89*
We do not yet know if we can provide epistemic characterizations using
such Kripke structures. 
%raf100: reordered
%raf90*: needs more here, i would like to argue that it might not be a bad thing that they don't exist, for instance in roshambo we might want ``chaos'' to be the right answer?
\item A natural next step would be to introduce notions of computation
in the epistemic logic.  
%raf44: maybe mention logical omniscience problem.
%joe26: 
There has already been some work in this direction (see, for example,
%joe57
%\cite{HMT,HMV94,Mos}).  We believe that, by combining the ideas of this
\cite{HMT,HMV94,Mos}).  We believe that combining the ideas of this
paper with those of the earlier papers will allow us to get, for
example, a cleaner knowledge-theoretic account of zero knowledge than
that given by Halpern, Moses, and Tuttle \citeyear{HMT}.
A first step in this direction is taken in \cite{HPR09}.

\item Finally, it would be interesting to use behavioral experiments to,
for example, determine the ``cost of computation'' in various games (such as
%joe85
%the finitely repeated prisoner's dilemma). 
finitely repeated prisoner's dilemma). 
\end{itemize}

\section{Acknowledgments}
%raf105: many of these are from both of us, but i am happy to rewrite
%anyway you prefer.
We wish to thank
%The second author wishes to thank 
Silvio Micali and abhi shelat for 
many exciting discussions about cryptography and game theory,
%raf62:
%raf64: should we add also Ehud?
%joe61: why not
%joe92
%Ehud
Ehud
%joe74
%and Adam Kalai for enligthening discussions.
and Adam Kalai for enlightening discussions,
%joe92
and Dongcai Shen for useful comments on the paper.
%joe56
%raf65: removed
%raf98: put back (it was probably removed for the crypto paper)
%joe92
%We also think Tim Roughgarden for encouraging us to think of conditions
We also thank Tim Roughgarden for encouraging us to think of conditions
that guarantee the existence of Nash equilibrium.

\bibliography{z,joe}
\newpage

\end{document}